\documentclass[aps,prd,preprint]{revtex4}
\usepackage{graphicx,amssymb,amsmath}

\begin{document}

\font\bbb=msbm10
\def \R {\hbox{\bbb R}}
\def \E {\hbox{\bbb E}}
\def \V {\hbox{\bbb V}}
\def \P {\hbox{\bbb P}}
\def \C {\hbox{\bbb C}}

\newcommand{\normal}[2]{{\rm Gaussian(\mbox{$#1|#2$})}}
\newcommand{\poisson}[2]{{\rm Poisson}(\mbox{$#1|#2$})}
\newcommand{\cov}[1]{{\rm Cov}(#1)}
\newcommand{\btheta}{\mathbf{\theta}}
\newcommand{\blambda}{\mathbf{\lambda}}
\newcommand{\bTheta}{\mathbf \Theta}
\newcommand{\bLambda}{\mathbf \Lambda}
\newcommand{\bomega}{\mathbf \omega}
\newcommand{\bSigma}{\mathbf \Sigma}
\newcommand{\htheta}{\hat \theta}
\newcommand{\hTheta}{\hat \Theta}
\newcommand{\thetaz}{\theta_{0}}
\newcommand{\lum}{{\mathcal L}}

\newcommand{\flatprior}[1]{\pi_{\rm F}(#1)}
\newcommand{\refprior}[1]{\pi_{\rm R}(#1)}
\newcommand{\prior}[1]{\pi(#1)}
\newcommand{\pdf}[2]{p(#1 | #2)}
\newcommand{\cprob}[2]{\textrm{Pr}(#1|#2)}
\newcommand{\prob}[1]{\textrm{Pr}(#1)}

\newcommand{\Equation}[1]{Equation (\ref{eq:#1})}
\newcommand{\Eq}[1]{Eq.\ (\ref{eq:#1})}
\newcommand{\Eqs}[2]{Eqs.\ (\ref{eq:#1}) and (\ref{eq:#2})}
\newcommand{\Eqns}[3]{Eqs.\ (\ref{eq:#1}), (\ref{eq:#2}) and (\ref{eq:#3})}
\newcommand{\Fig}[1]{Fig.\ \ref{fig:#1}}
\newcommand{\Figure}[1]{Figure\ (\ref{fig:#1})}
\newcommand{\Sec}[1]{Sect.\ \ref{sec:#1}}

\def\met{\mbox{${\hbox{$E$\kern-0.6em\lower-.1ex\hbox{/}}}_T$}}

\preprint{
    \vbox{
        \rightline{Version 1.8}\break
   }
}

\title{Priors for New Physics}

\author{Maurizio Pierini$^1$, Harrison B. Prosper$^2$, Sezen Sekmen$^2$ and
Maria Spiropulu$^{1,3}$}

\affiliation{$^1$CERN, CH-1211, Geneva 23, Switzerland}

\affiliation{$^2$Department of Physics, Florida State University, Tallahassee, Florida 32306, USA}

\affiliation{$^3$Department of Physics, Caltech, Pasadena, California 91125, USA}

\date{\today}

\begin{abstract}
The interpretation of data in terms of 
multi-parameter models of new physics, using the Bayesian approach, requires  the construction of multi-parameter
priors. We propose a construction that uses elements of Bayesian reference analysis. 
Our idea is to initiate the chain of inference with the reference prior for a 
likelihood function that depends on a single parameter of interest that is
a function of the parameters of the physics model. 
The reference posterior density of the parameter of interest induces on the 
parameter space of the physics model a
\emph{class} of posterior densities. We propose to continue the chain of inference with a particular density from this class, namely, the one for which indistinguishable models are equiprobable and use it
 as the prior  for subsequent analysis. We illustrate our method by 
applying it to the constrained minimal supersymmetric Standard Model and two 
non-universal variants of it.
\end{abstract}

\pacs{}

\maketitle

\section{Introduction}
\label{sec:introduction}
With the start of the Large Hadron Collider (LHC)~\cite{LHC}, we have entered an era in which
speculation about new physics has given way to detailed experimental study. This
has had the welcome consequence of focusing attention on a difficult practical 
question: given the plethora of models of potential new physics, many depending on 
multiple unknown parameters, what is the best practical way to navigate the landscape of
possibilities? This is a multi-faceted problem, of which undoubtedly the most challenging is
devising reliable background estimates for all the final states that are being scrutinized. Another
challenge is the construction of very fast accurate simulations~\cite{Delphes} of new physics models at hundreds of thousands, even millions, of parameter points. This is necessary because, in general, the effective cross section, $\epsilon(\theta) \sigma(\theta)$---that is, the signal efficiency, $\epsilon(\theta)$, times cross section, $\sigma(\theta)$---is a function of the parameters $\theta$ of the model under investigation.

In this Paper, we shall assume that both of these difficult tasks have been accomplished.
Instead we address another important facet of the problem,
namely, that of extracting information about a given new physics model once
LHC data become sufficiently abundant to test it. 
We propose a new method that is applicable to
any multi-parameter model that yields a prediction about the expected signal
count. We illustrate the method using three supersymmetric (SUSY) models~\cite{SUSY}: the
constrained minimal supersymmetric Standard Model (CMSSM)~\cite{cmssm} 
and two non-universal variants of it.

The availability of  increasingly powerful computers has made it possible
 to study multi-parameter models in a holistic manner.  Indeed, it has become 
 routine to use techniques such
as Markov Chain Monte Carlo (MCMC)~\cite{MCMC} to explore the multi-dimensional
parameter spaces of models such as SUSY~\cite{SUSYMCMC}. This is another welcome development.
 Recent work on SUSY models~\cite{pMSSM} 
 has shown that a holistic
approach can yield qualitatively different conclusions from those arrived
at using the traditional approach based on benchmarks~\cite{CMSbenchmarks}. SUSY models have been studied using both frequentist~\cite{SUSYfreq}
and Bayesian~\cite{SUSYBayes} methods. The frequentist studies typically construct confidence regions and obtain the best-fit point. Sometimes, information about individual parameters or pairs of parameters is obtained by
projecting the likelihood function onto the parameters of interest. This procedure is actually a frequentist/Bayesian
hybrid, which amounts to using a flat prior
on the parameters. A conceptually more consistent, albeit approximate, frequentist approach is to 
construct a profile likelihood~\cite{Cowan:2010js,Feroz:2011bj,Akrami:2009hp} for the parameter of interest. For example, if
the parameter of interest is $m_0$ and $l(m_0, \omega) \sim p(x | m_0, \omega)$ is the likelihood
function for observations $x$, where  $\omega$ denotes the remaining parameters, the
profile likelihood for $m_0$ is $l_P(m_0) \sim p(x|m_0, \hat{\omega}(m_0))$, where $\hat{\omega}(m_0)$ is
the best fit value of the parameters $\omega$ for a given value of $m_0$. The profile
likelihood $l_P(m_0)$ is then used
as if it were a true likelihood. 

We propose to use the Bayesian approach~\cite{Bayes} because of its strong theoretical foundations,
its generality and the fact that it is conceptually straightforward: given a prior $\pi(\theta)$ 
defined on the parameter space $\Theta$ of the model, where in general $\theta$ 
is multi-dimensional, and a likelihood $p(x|\theta)$, one
computes the posterior density $p(\theta|x) \sim p(x|\theta) \, \pi(\theta)$ from which a myriad of
details can be extracted such as point estimates or credible 
regions. It is also possible to make predictions about which data would be most useful to take next, and
one can rank models according to their concordance with observations. Moreover, all manner of uncertainties, irrespective of their provenance and how we choose to  label them---statistical, systematic, theoretical, best guess, etc.---can be accounted for in a conceptually coherent 
and unified manner.

Every fully
Bayesian analysis, however, must contend with the problem of 
constructing 
a prior $\pi(\theta)$ on the parameter space of the model under
investigation. This task is especially difficult in circumstances in which intuition provides little
guidance as is invariably the case for multi-parameter models.  Current studies, which place
flat or logarithmic priors on the parameters of new physics models, are sensitive to the choice of prior~\cite{SUSYBayes}; therefore, the choice of prior is a critical issue that must be squarely faced.
This is  the main purpose of this Paper.

The current sensitivity of results to the prior is sometimes construed as an intrinsic difficulty with
the Bayesian approach. In fact, the correct conclusion to be drawn is
that it is not yet possible to place robust constraints on all the parameters of a typical multi-parameter model of new physics, a conclusion that is independent of the method used to extract information about the model be it frequentist or Bayesian. 
The difficulty is not that results are sensitive to the prior---this fact tells us something obvious and important:
we need more data and better analyses. Rather the difficulty is that 
flat priors on multi-dimensional parameter spaces can lead to pathological results,
which may not be apparent without a careful study.  Flat priors have 
been used successfully, witness the recent discovery of single top quark
production by D\O~\cite{D0singletop} and CDF~\cite{CDFsingletop}. But these results
were obtained with a flat prior applied to a \emph{single} carefully chosen parameter, namely, the 
cross section~\cite{Bertram2000}. 

Given that
our multi-dimensional intuition may be unreliable, we are faced with a choice: either abandon the Bayesian
approach---and, in our view, abandon an extremely powerful set of ideas---or, as we propose, 
put intuition aside and use a formal 
procedure with mathematically verifiable properties to place priors
on the parameter spaces.  We propose
a solution inspired by  a set of Bayesian methods called
\emph{reference  analysis}~\cite{Bernardo, Demortier2005, Demortier2010}, whose
key construct is the \emph{reference prior}.
 
We advocate the use of
reference priors because they lead to inferences with useful properties, including invariance
under one-to-one transformations of the parameters and excellent frequentist coverage. The
latter property means that the (Bayesian) credible regions are also approximate  (frequentist) confidence regions. Moreover, the reference prior can be perturbed in a controlled way to check the 
robustness of conclusions.

Having initiated the inference chain with a
reference prior, we can use Bayesian methods to
\begin{itemize}
\item quantify the statistical significance of a signal,
\item rank models according to their concordance with observations,
\item estimate model parameters, and
\item design an optimal analysis for a given model and a given integrated luminosity.
\end{itemize}
In this Paper, in addition to the main task of constructing multi-parameter priors, we address 
the first two points---the statistical significance of a signal and model ranking---and we
defer consideration of the last two to a future publication.  

Bayesian
reference analysis~\cite{Bernardo, Demortier2005, Demortier2010} provides a principled
way to approach the problem of multi-parameter priors. However, while the solution it proposes is computationally
feasible for one-parameter problems, it rapidly becomes computationally prohibitive for
multi-parameter problems using current algorithms. Since the 1-parameter problem is a well-understood, solved, problem, our proposed solution  begins
with the solution of a 1-parameter problem and 
proceeds to the
multi-parameter problem by imposing two requirements on the multi-parameter prior: consistency and 
equiprobability, both of which are described in detail below.

Our solution
proceeds in four steps:
\begin{enumerate}
\item first, we compute the marginal likelihood by integrating the likelihood function
with respect to an evidence-based prior over
all parameters except the parameter of interest;
\item next, we compute the reference prior associated with the marginal likelihood;
\item then, we compute the reference posterior
density for the parameter of interest,
\item and, finally, we map the reference posterior density to a posterior density on the parameter space of each
multi-parameter model under study.
\end{enumerate}
Clearly, these steps can be applied to any experiment that has a single parameter of interest.
In this Paper, we apply the steps to a single count experiment because it yields the simplest possible analysis and the key calculations can be done exactly. In the following sections, we describe the single count model, its reference prior, and our
method for mapping the signal posterior density to the parameter space of a given
multi-parameter model.

The Paper is organized as follows. In Sec.~\ref{sec:singlecountmodel}, we give a detailed description of the single count model and its associated reference prior. Our construction of multi-parameter priors is described in Sec.~\ref{sec:method}. In Sec.~\ref{sec:examples}
we illustrate the method using three SUSY models,
a 2-parameter CMSSM and 
two 5-parameter non-universal generalizations.
We end with a summary and concluding remarks.

\section{The Single Count Model}
\label{sec:singlecountmodel}
In the context of the LHC, the single count model describes the results of a ``cut and count" 
analysis in which
$N$
proton-proton collision events are found to pass a given set of selection criteria, that is, cuts. The expected 
number of events, $n$, is given by
\begin{equation}
n = \mu + s,
\label{eq:n}
\end{equation}
where $\mu$ is the expected number of Standard Model background events and $s \geq 0$---assumed to be purely additive---is the expected 
number of signal events due to
(unknown) new physics. The observed count is denoted by $N$ and the expected (that is, mean) count is denoted by $n$. We shall use upper case letters
for observed values and lower case letters for expected values. 

The result of \emph{any} experiment can be encoded in its likelihood
function, the probability density function (pdf) of the observations (sometimes
called the probability mass function if the data are
discrete) evaluated at the actual observations. From the likelihood
function and the prior density for the expected signal and background we can 
compute the posterior probability $\cprob{s}{N}  = \pdf{s}{N} \, ds$ of the signal, that is, 
the probability that the expected signal lies in the interval $\delta = (s, s + ds)$, given the observed count
$N$.

We choose to parametrize the likelihood in terms of the
expected signal
$s$ rather than the cross section $\sigma$, as is done in Ref.~\cite{Demortier2010}, 
so that the results of the counting experiment remain independent of the new physics model. The cuts may have been motivated by a specific model of new physics, however,
the signal posterior density can be interpreted using \emph{any} physics model that makes
predictions for the expected signal in the final states considered. Moreover,
as we shall see,  we can devise a purely Bayesian measure of the  degree to which the observation of $N$ events favors
the
hypothesis $s > 0$ rather than the background-only hypothesis $s = 0$, independently
of any presumed model of new physics. Moreover, this can be readily generalized to a
multi-count analysis.

For a counting experiment that yields $N$ events, we make the usual assumption that the likelihood function is
given by a Poisson distribution,
\begin{align}
p(N | \mu, s) &\;=\; \poisson{N}{\mu + s}, 
\label{eq:likelihood}
\end{align}
 with mean $\mu + s$. The associated 2-parameter 
prior, $\pi(\mu, s)$, can be factorized in two ways,
\begin{align}
\label{eq:method1}
\pi(\mu, s) & =  \pi(s | \mu) \; \pi(\mu), & \quad\quad\textrm{Method 1 }\\
\label{eq:method2}
\pi(\mu, s) & =  \pi(\mu | s) \; \pi(s), 	& \quad\quad\textrm{Method 2},
\end{align}
both of which were considered in Ref.~\cite{Demortier2010}. Here, we
consider Method 2 only. We do so because we can reduce the likelihood function $p(N|\mu, s)$
to a function of the single parameter $s$ through marginalization,
\begin{equation}
p(N | s) = \int_0^{\infty} p(N | \mu, s ) \, \pi(\mu| s) \, d\mu,
\end{equation}
which permits the application of the 1-parameter reference prior
algorithm~\cite{Demortier2010} to compute the reference prior for the expected signal, while
avoiding the technical issue of nested compact sets~\cite{Demortier2010}.

Following Ref.~\cite{Demortier2010}, we
model the evidence-based prior 
$\pi(\mu|s)$
for the expected background 
by a gamma density,
\begin{align}
\label{eq:mu}
\pi(\mu|s) = \pi(\mu) &\;=\; \frac{b(b\mu)^{Y-1/2}}{\Gamma(Y+1/2)}\;
                    e^{-b\mu},
\end{align}
where $b$ and $Y$ are known constants. We further assume that the prior is independent of the expected
signal, $s$. (See Appendix~\ref{app:backgroundPrior} for its derivation.) Then, we  
integrate over $\mu$ to arrive at the 1-parameter marginal likelihood,
\begin{align}
p(N\,|\,s) 
            &\;=\;\int p(N\,| \, \mu, s)\;
                  \pi(\mu)\,d\mu,\nonumber\\[2mm]
            &\;=\;\int \frac{(\mu + s)^{N}}{N!}\;
                  e^{-\mu - s}\;
                  \frac{b(b\mu)^{Y-1/2}}{\Gamma(Y+1/2)}\;
                  e^{-b\mu}\;d\mu,\nonumber\\[2mm]
            &\;=
                  \left[\frac{b}{b+1}\right]^{Y+\frac{1}{2}}\;
                  \sum_{k=0}^{N}v_{Nk}\, \textrm{Poisson}(k|s),\nonumber\\[2mm]
\mbox{where}\quad 
v_{ik} &\;\equiv\;\frac{\Gamma(Y+\frac{1}{2}+i-k)}{\Gamma(Y+\frac{1}{2}) \; (i-k)!}\;
             \left[\frac{1}{b+1}\right]^{i-k},
\label{eq:MarginalModel}
\end{align}
for the expected signal, $s$, whose
reference prior, $\pi(s)$, is calculated in the next section.

\subsection{Reference Priors}
When we know almost nothing 
about a potential signal it seems prudent to use
a prior for the expected signal that is as
noncommittal as possible.
The approach in high energy physics has been to use a flat prior~\cite{Bertram2000} for a 
parameter about which little is known, or for which one wishes to act as if that is the case. 
But, for multi-parameter models,
our intuition is ill-equipped to choose the parameterization in terms of which the prior
is flat. 
We therefore propose a different approach. Our idea is to construct a prior  for each new physics model starting with the reference prior for an experiment with a single parameter of 
interest---here the expected signal, $s$, for a single count experiment.   
 By construction, a reference prior~\cite{Bernardo, Demortier2005, Demortier2010},  
 on average and given unlimited data, 
 maximizes the influence of the data relative to the prior. 
  
The
intuition that underlies the construction of such priors is that the influence
of the observations will be greatest  if the ``separation" between 
the posterior density  and the prior  is as large as possible. 
Reference analysis~\cite{Berger2009} quantifies the separation 
between the two densities $p(s | N)$ and $\pi(s)$ using the Kullback-Leibler (KL), 
divergence, which for the particular problem we address is given by
\begin{equation}
D[\pi,  p] \equiv \int \,  p(s | N)\;\ln\frac{p(s| N)}{\pi(s)} \, ds.
\label{eq:KL}
\end{equation}
This non-negative quantity, which is  invariant under one-to-one transformations of $s$ 
and zero if and only if the densities $p(s | N)$ and $\pi(s)$ are identical, may also be interpreted as a measure of the information gained from the (single count)
experiment.

Since we wish to maximize the influence of the observations, we might be tempted
to maximize Eq.~(\ref{eq:KL}) with respect to the prior, $\pi(s)$. This, however, would be
unsatisfactory because the prior would then depend on the specific observations, which 
would enter the
posterior density
twice: once in the prior and once in the likelihood. It is more satisfactory to use the
average of $D[\pi, p]$
over all possible observations. Integration over the space of observations---standard
practice in the frequentist approach---may seem
a decidedly un-Bayesian thing to do. However, the likelihood principle~\cite{LP}, the idea
that inferences should be based on the observed data only, makes sense only if
we actually have observations. Obviously, before we perform the analysis, we do not know the 
value of the count $N$; therefore, since the count is unknown we should average over all possible realizations of $N$. Once we know the count, our  inferences should be based on $N$ only. 
For completeness, we give the key details of the reference prior algorithm
in the Appendix~\ref{app:refPrior}. 

The calculation of reference priors simplifies considerably for posterior densities
that  are asymptotically normal, that
is, that become Gaussian as more and more data are included. In this case, the reference prior coincides with the Jeffreys' prior~\cite{Berger2009},
\begin{equation}
\pi(s) = \sqrt{\mathbb{E}\left[ -\frac{d^2 \ln p(N | s)}{ds^2}\right]},
\label{eq:JeffreysPrior}
\end{equation}
where for the single count model the expectation is with respect to the  (marginal) 
likelihood $p(N | s)$, given in Eq.~(\ref{eq:MarginalModel}).
For a counting experiment, the asymptotic form of the posterior density $p(s | N)$ 
is indeed Gaussian. Therefore, the
reference prior for $p(N | s)$ can be computed using  Eq.~(\ref{eq:JeffreysPrior}). 
Adapting the results of Ref.~\cite{Demortier2010}, we find,
\begin{align}
\label{eq:piR2p}
\pi(s) &\;=\; \sqrt{ \sum_{i=0}^{\infty}
\frac{[T_{i}^{0}(s)\,-\,T_{i}^{1}(s)/s]^{2}}{T_{i}^{0}(s)}},\nonumber\\[2mm]
\mbox{where}\quad
T_{i}^{m}(s) & \;\equiv\;\sum_{k=0}^{i} k^{m}\;v_{ik}\;
\textrm{Poisson}(k|s)\quad\textrm{for}\quad m=0,1,
\end{align}
and $v_{ik}$ are the coefficients defined in Eq.~(\ref{eq:MarginalModel}).
The complete reference prior, $\pi(\mu, s)$, is the product of Eqs.~(\ref{eq:mu}) and (\ref{eq:piR2p}),
while the complete reference posterior density is 
\begin{equation}
p(\mu, s | N) = \frac{p(N | \mu, s) \, \pi(\mu, s)} 
{\int_0^{\infty} ds \int_0^{\infty} d\mu \, p(N | \mu, s) \, \pi(\mu, s)}.
\end{equation}
The reference posterior density for the expected signal is obtained by integrating over $\mu$,
\begin{eqnarray}
p(s | N ) & = & \int_0^{\infty} p(\mu, s | N) \, d\mu, \nonumber \\
 & = & p(N | s) \, \pi(s) / \int_0^{\infty} p(N | s ) \, \pi(s) \, ds,
\label{eq:psN}
\end{eqnarray}
where $p(N | s)$ and $\pi(s)$ are given by Eqs.~(\ref{eq:MarginalModel}) and (\ref{eq:piR2p}), respectively. (See Appendix~\ref{app:likelihood} for more technical details.)

\subsection{A Measure of Signal Significance}
Assessing the statistical significance of a signal is a standard analysis task in 
high energy physics~\cite{Cousins:2008zz}, one which 
traditionally has been done with a $p$-value~\cite{Cowan:2010js}. Here we propose an alternative
measure that uses the reference posterior density $p(\mu, s | N)$.

Suppose we are
given some function $\delta(\mu, s)$ that measures the separation between
the (composite) background plus signal hypothesis,  $H_1: \mu > 0, s > 0$, and the (composite) background-only hypothesis, 
$H_0 : \mu > 0, s = 0$.  If the separation between the hypotheses were large enough
then presumably we would 
reject the background-only hypothesis in favor of the alternative.  But, since we
know neither the expected background $\mu$ nor the expected signal $s$,
the natural Bayesian
thing to do is to average $\delta(\mu, s)$
with respect to all possible hypotheses about the values of $\mu$ and $s$,  
\begin{eqnarray}
d(N) \equiv \mathbb{E}[ \delta(\mu, s)] & = &  \int_0^{\infty} ds \int_0^{\infty} 
d\mu \, \delta(\mu, s) \; p(\mu, s | N), \nonumber \\
	& = & \int_0^{\infty} ds \int_0^{\infty} 
d\mu \, \delta(\mu, s) \; p(N| \mu, s ) \; \pi(\mu, s) / p(N),
\label{eq:dN}
\end{eqnarray}
where $p(N)$ is the normalization constant $p(N) =  \int_0^{\infty} ds \int_0^{\infty} 
d\mu \, p(N| \mu, s ) \; \pi(\mu, s)$. If $\delta(\mu, s)$ is interpreted as a loss function then $d(N)$ is a measure of the loss incurred, on average, 
if one were to stubbornly adhere to the background-only hypothesis regardless of the outcome of
the experiment.
A signal is declared to be statistically significant if
 $d(N) > d^*$, where $d^*$ is some agreed-upon
threshold. Moreover,  the decision to accept or reject $H_0$ and 
thereby reject or accept the
alternative $H_1$ may be taken independently of any model of
new physics. 

There are many possible choices for the function $\delta(\mu, s)$.  We
propose to use the
      Kullback-Leibler divergence~\cite{Bernardo, Demortier2005}, 
\begin{eqnarray}
      \delta(\mu, s) & \;=\; & \sum_{k=0}^{\infty} p(k\,| \, \mu + s) 
                    \ln\frac{p(k\,|\, \mu + s)}{p(k\,|\,\mu)}, \nonumber \\
                    & \;=\; & -s + (\mu+s) \ln(1 + s/\mu),
                    \label{eq:loss}
\end{eqnarray}
      between the densities 
      $p(k | \mu + s)$ and $p(k | \mu)$ associated with hypotheses $H_1$ and $H_0$,
            respectively.
For fully specified models, Eq.~(\ref{eq:loss}) is simply the expected 
      log-likelihood ratio. We can gain some insight into $\delta(\mu, s)$ by
      considering a counting experiment for which  $s << \mu$, which
      characterizes early searches for new physics. In this limit~\footnote{In this limit---essentially, when the two hypotheses $H_1$ and $H_0$
are nearly degenerate---the KL divergence can
be interpreted as twice the square of the
distance between the associated densities in the space of functions~\cite{metric}.},
      \begin{equation}
      \delta(\mu, s) = -s+(s+\mu)\ln\Bigl(1+\frac{s}{\mu}\Bigr)
                            \approx -s+(s+\mu)\Bigl[\frac{s}{\mu} -
                                   \frac{1}{2}\frac{s^2}{\mu^2}+\cdots\Bigr]
                            \approx \frac{1}{2} \frac{s^2}{\mu},
      \end{equation}
      that is, $\sqrt{2 \; \delta(\mu, s)} \sim s / \sqrt{\mu}$.
This suggests taking the quantity, 
      \begin{equation}
      q \equiv \sqrt{2 \; d(N)},
      \end{equation}
as a Bayesian analog of the well-known (and oft-abused)
measure of ``signal significance," $q = s / \sqrt{\mu}$. As such, it is an analog of an
``n-sigma," that is, the standard
re-scaling of a $p$-value using the single tail area of a normal density~\cite{Cowan:2010js}. This
\emph{approximate} correspondence
provides a simple calibration of $d(N)$.

\subsubsection{Generalization to Multiple Counts}
For an experiment that yields $K$ independent counts, $N_k$, $k = 1, \cdots, K$, with expected
background and signal counts $\mu_k$ and $s_k$, respectively, the
KL divergence is simply the sum
\begin{equation}
	\delta(\mu_1,s_1,\cdots) = \sum_{k=1}^K \delta(\mu_k, s_k),
\end{equation}
over terms $\delta(\mu_k, s_k)$, each of which is given by \Eq{loss}, while the
signal significance measure generalizes to
\begin{eqnarray}
d(N_1,\cdots) 	& \equiv & \mathbb{E}[ \delta(\mu_1, s_1,\cdots)] \nonumber \\
			& =  & \int_0^{\infty} ds_1 \int_0^{\infty} 
d\mu_1 \cdots  \int_0^{\infty} ds_K \int_0^{\infty} d\mu_K \, \delta(\mu_1, s_1, \cdots) \; \nonumber \\
			& \times & p(\mu_1, s_1, \cdots | N_1, \cdots), \nonumber \\ \nonumber \\
& =  & \sum_{k=1}^K \int_0^{\infty} ds_1 \int_0^{\infty} 
d\mu_1 \cdots  \int_0^{\infty} ds_K \int_0^{\infty} d\mu_K \, \delta(\mu_k, s_k) \; \nonumber \\
			& \times & p(N_1| \mu_1, s_1) \; \pi(\mu_1, s_1) / p(N_1) \cdots 
			p(N_K| \mu_K, s_K) \; \pi(\mu_K, s_K) / p(N_K), \nonumber \\
			& = & \sum_{k=1}^K d(N_k),
\label{eq:dN1NK}
\end{eqnarray}
where we have used the fact that the posterior density $p(\mu_1, s_1, \cdots | N_1, \cdots)$
factorizes into a product of $K$ terms, one for each count $N_k$, each of which integrates
to one.

%
%

\section{Multi-Parameter Priors and Model Ranking}
\label{sec:method}

\subsection{Multi-Parameter Priors}
We have a well-defined reference posterior density for the signal, $p(s|N)$, which satisfies
\begin{equation}
\int_0^{\infty} ds \, p(s|N) = 1.
\label{eq:psNnorm} 
\end{equation}
Our 
task now is to map it to a density $p(\theta)$  on the parameter
space $\Theta$ of a given physics model. 

By assumption, the
model predicts the expected
signal $s$ via a predictor function $s = f(\theta)$.
Consequently, the reference posterior density $p(s|N)$
 induces, or is consistent with, posterior densities on $\Theta$ that satisfy~\cite{Gillespie}
 \begin{eqnarray}
p(s | N)  & = & \int_{\Theta} \delta[s - f(\theta)] \, p(\theta) \, d\theta.
		\label{eq:intmap}
\end{eqnarray}
Equation~(\ref{eq:intmap}) is the consistency requirement we alluded to. Note, \Eqs{psNnorm}{intmap} imply that $\int d\theta \; p(\theta) = 1$.

\Equation{intmap} determines $p(\theta)$ only to within a class. Therefore, 
 we need a plausible way to choose a specific function from that class that would serve
 as a suitable posterior density and hence a prior for subsequent analysis. 
To that end, we note that
every point $\theta\in\Delta$, where $\Delta$ is the image of $\delta = (s, s + ds) \in \R$, is associated
with the \emph{same} expected signal $s \in \delta$.
In that sense, the points in $\Delta$ are
indistinguishable; that is, $\Delta$ defines a set of ``look-alike'' (LL)
models. We therefore propose that $p(\theta)$ be chosen so that
\begin{equation}
 \mbox{\emph{every point within
$\Delta$ is equiprobable},}
\label{eq:lookalike}
\end{equation}
that is, that the density $p(\theta)$ be constant over $\Delta$.
This choice yields the following expression for 
$p(\theta)$,
\begin{equation}
p(\theta) = p(s(\theta) | N) \, / \, A(s(\theta)), 
\label{eq:ptheta}
\end{equation}
where,
\begin{equation}
A(s) =  \int_{\Theta} \delta[s - f(\theta)] \, d\theta,
\label{eq:Area}
\end{equation}
is the \emph{area} of the hyper-surface defined by
$s - f(\theta) = 0$. This choice is arguably the simplest  for
$p(\theta)$ given that the \emph{only} information at hand is the reference posterior density for the signal. If, however, one has cogent information about how $p(\theta)$ should vary on these
hyper-surfaces, then our simple choice can be replaced with something
consistent with this information and \Eq{intmap}.

There are two technical challenges in our proposed method. The first is that, in general, we do not 
have explicit functional forms for the mapping $s = f(\theta)$. In practice, in order to calculate 
the expected signal, we simulate a large number of signal events for a given parameter point $\theta$, we apply cuts to these events and we determine what fraction of them survive
the cuts; that is, we calculate the signal efficiency $\epsilon(\theta)$. Then, for a given
integrated luminosity ${\cal L}$ , we compute the expected signal using $s = \epsilon(\theta)\,\sigma(\theta) \, {\cal L} \equiv f(\theta)$, where $\sigma(\theta)$ is the cross section.  The second challenge is the calculation of the surface term, \Eq{Area}. We discuss both of these calculations
in Sect.~\ref{sec:examples}, in which we illustrate the practical application of our method. But
first we briefly review the standard Bayesian approach to model ranking. 

\subsection{Model Ranking}
\label{sec:modelranking}
If Nature is kind to us, we shall eventually start to see signals of new physics at the LHC. Then, the most important tasks will be to characterize the observations experimentally and determine which candidate model best describes them.

Suppose we wish to rank $M = 1,\cdots, J$ candidate models of new physics according to 
their concordance with the observations. In general, each model will have its own set of parameters 
$\theta_M$, perhaps differing in meaning and, or, dimensionality. The standard Bayesian approach to model ranking is, as usual, direct: calculate the probability of each model $M$~\cite{ModelRanking}
given the observations. The model with the highest probability wins.

Given the likelihood function $p(\textrm{data}|\theta_M, M)$ and prior $\pi(\theta_M, M) = \pi(\theta_M|M) \; \pi(M)$, we first compute the \emph{evidence}~\cite{ModelRanking},
\begin{equation}
	p(\textrm{data}|M) = \int d\theta_M \; p(\textrm{data}|\theta_M, M) \; \pi(\theta_M|M),
	\label{eq:evidence}
\end{equation}
and then the probability of each model 
\begin{equation}
	P(M|\textrm{data}) = p(\textrm{data}| M) \; \pi(M) / \sum_{M=1}^J p(\textrm{data}| M) \; \pi(M),
	\label{eq:probM}
\end{equation}
where $\pi(M)$ is a discrete prior probability distribution over the space of models. The polemical aspect of \Eq{probM} is the need to specify the values of $\pi(M)$, on which there seems little chance of agreement.  If, however, the models are
judged to be equally implausible---or if the LHC experiments were to reach an accord to that effect, it would be appropriate to set $\pi(M) = 1 / M$, in which case
\Eq{probM} reduces to
\begin{equation}
	P(M|\textrm{data}) = p(\textrm{data}| M)  / \sum_{M=1}^J p(\textrm{data}| M).
	\label{eq:probM2}
\end{equation}
Absent such an accord, it is still possible to rank models using their evidences: the larger the evidence the more favored is the model. 

But, there is an important caveat: it is necessary to use proper priors for $\pi(\theta_M|M)$, that is, priors that integrate to one. An improper prior is defined only to within an arbitrary scale factor. Consequently, were such a prior to be used to compute the evidence, the latter would be defined only to within the same arbitrary scale factor. Therefore, in order for the evidences to be well-defined, the priors must be proper. By
construction, this is the case for the multi-dimensional priors introduced above.

Models can also be ranked using Bayesian reference analysis. However, we defer the discussion
to a future publication.

\section{Illustrative Examples}
\label{sec:examples}
Our proposed method for constructing multi-parameter priors is quite general. It can be applied, in principle, to any physics model of any dimensionality provided that the  model makes
a prediction for the parameter of interest, which in our case is
the expected signal in a counting experiment. For simplicity, however, we illustrate the
application of the method  using a SUSY model with only two free parameters for which
the results are easily visualized. We then consider two 5-parameter models.

\subsection{2-D Model}
\label{sec:2dExample}
The first model we consider is the sub-model of the CMSSM~\cite{cmssm} defined by
the free parameters $m_0$,
$m_{1/2}$, and the fixed parameters $\tan\beta=10$, $A_0=0$ and $\mu > 0$.  We
take the CMS  benchmark
point LM1~\cite{CMSbenchmarks}, defined by the 
fixed parameters $m_0=60$,
$m_{1/2}=250$, $\tan\beta=10$, $A_0=0$ and $\mu > 0$, as our \emph{true state of Nature} (TSN), which
provides the ``observed" count $N$~\cite{Pierini:2011yf}.  For each point in a grid of points in the $m_0 - m_{1/2}$
plane, including the point LM1, the SUSY spectrum is calculated using {\tt SOFTSUSY 3.1}~\cite{softsusy} and sparticle decays using {\tt SUSYHIT}~\cite{susyhit}. We generate 1000 7 TeV LHC events using {\tt PYTHIA 6.4}~\cite{Pythia} and approximate the 
response of the CMS detector~\cite{CMSdetector} to these events using a modified version of the fast detector simulation program {\tt PGS}~\cite{PGS}. We apply a CMS multijets plus missing
transverse energy ($\met$) event selection~\cite{CMSmultijetsMET} to the events simulated at
each point $\theta = (m_0, m_{1/2})$ and we take the background estimates from the
CMS analysis in Ref.~\cite{CMSmultijetsMET}.

Three hypothetical results are considered: i)
$N=3$ events observed in ${\cal L} = 1$ pb$^{-1}$ of data; ii) $N=270$ events observed in
100 pb$^{-1}$, and $N=1335$ events observed in 500 pb$^{-1}$. In each case,
we compute the posterior density $p(s | N)$ for the expected signal
count at each point in the $m_0 - m_{1/2}$ plane and map it to the posterior density $p(m_0, m_{1/2})$, which we take as the
prior $\pi(m_0, m_{1/2})$. The value of the surface term in this case is simply the length of the curve
$s - f(m_0, m_{1/2}) = 0$.

\begin{figure}[hbtp]
  \begin{center}
    \includegraphics[width=1\textwidth]{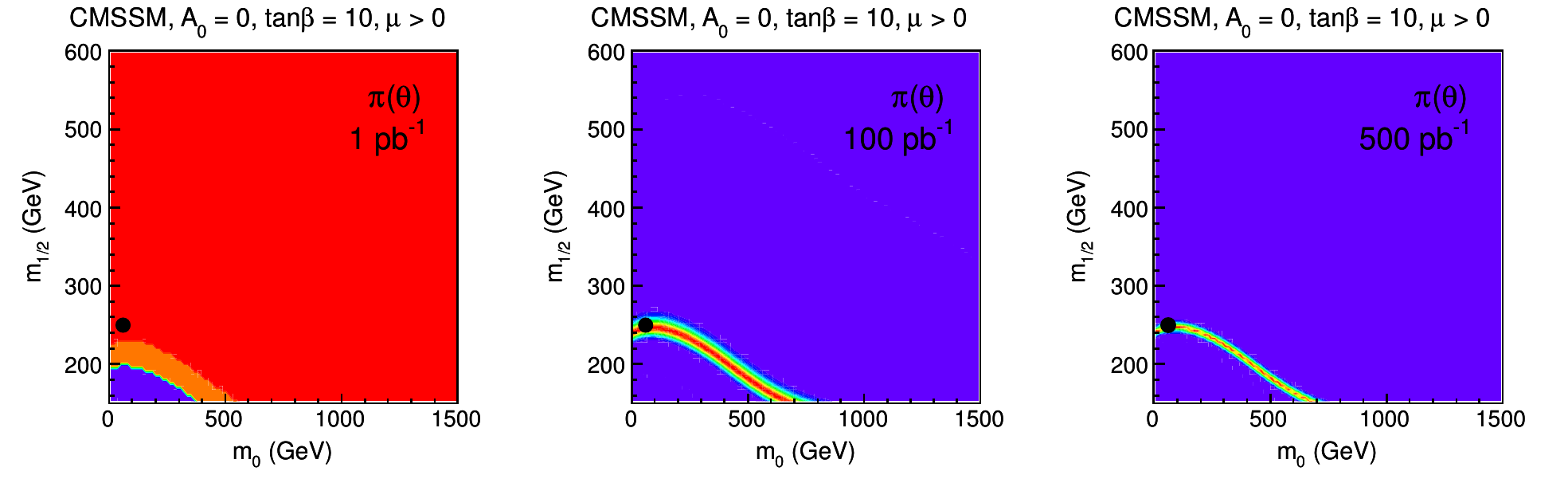}
    \caption{Induced posterior densities on the $m_0 - m_{1/2}$ plane for 1
      pb$^{-1}$ (left), 100 pb$^{-1}$ (center), and 500 pb$^{-1}$
      (right).  The TSN is indicated by the black dot. 
      \label{fig:priorsLM1}}
  \end{center}
\end{figure}

The plots in Fig.~\ref{fig:priorsLM1} show the induced posterior density $p(m_0, m_{1/2})$,
and hence prior $\pi(m_0,m_{1/2})$, for the three integrated luminosities.
The plots show several nice
features. For low statistics, the prior
is featureless in the region to which the experiment has no
sensitivity, while the low mass region is disfavored. At moderate
luminosity the prior peaks at the right value, favoring the correct
model and, with the same probability, all its LL models. At large
luminosity the prior converges to the correct LL sub-space $\Delta$, which, as noted, is
a curve.  

The fact that the sub-space is not a single point shows that an infinite amount of data does not
necessarily guarantee the irrelevance of the prior that initiated the chain of inference. This is
why choosing the prior carefully is important.
Since the LL sub-space $\Delta$ is extended, it remains sensitive to the initiating prior, which because
of the manner in which we choose
to  map $p(s|N)$ to $p(m_0, m_{1/2})$ is constant across the LL sub-space. The upshot of this is that we
should expect the initiating
prior to become irrelevant only if an analysis is able
to break the model degenaracy  so that with an infinite amount of data the
LL sub-space 
collapses to a point or, more realistically, to a very small sub-space over which the variation of the initiating prior
is negligible.

The
degeneracy between models with the same expected signal count---which
we argue is a desirable property---is intrinsic to
the approach we propose. However, having defined a prior over
the parameter space
of the model under study, we can move well beyond a simple counting experiment.
SUSY models have the virtue of making numerous predictions that can be tested
in a variety of ways. We argue that the interpretation of data at the LHC should be done
in a manner that is consistent with all the  tested predictions of the model under
consideration. To do otherwise risks reaching scientifically untenable conclusions: for example, 
that a region of parameter space is still allowed when a more complete analysis 
might say quite the opposite. If we have access to results from different analyses, perhaps from different
experiments, we argue that a consistent analysis should incorprate these results whenever possible. 
The ability to do this in a systematic manner is one of our motivations for addressing the problem of multi-parameter
priors. 

\begin{figure}[hbtp]
  \begin{center}
    \includegraphics[width=1\textwidth]{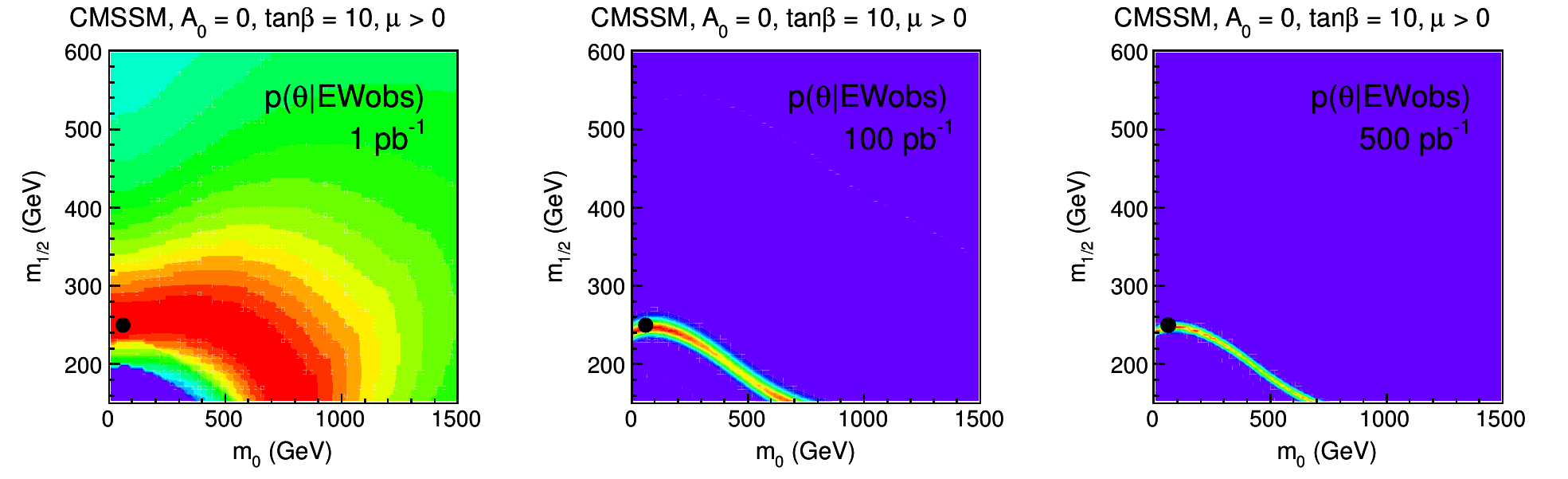}
    \caption{Posterior density induced on the $m_0 - m_{1/2}$ plane, after the inclusion of the electroweak observables, for 1
      pb$^{-1}$ (left), 100 pb$^{-1}$ (center), and 500 pb$^{-1}$
      (right). The TSN is indicated by the black dot.  The central values of the electroweak observables are computed at the TSN point, but we use the experimental uncertainties from Refs.~\cite{pdg}. \label{fig:posteriorLM1}}
  \end{center}
\end{figure}

In order to break the model degeneracy, we can incorporate the likelihood associated with
a set of additional observables $\vec x$ and compute the
posterior density $p(m_0, m_{1/2}|\vec x)$ using the prior $\pi(m_0, m_{1/2})$ computed
from the single count analysis. An example is given in
Fig.~\ref{fig:posteriorLM1}, where the function,
\begin{equation}
p(m_0, m_{1/2}|\vec x) \propto p(\vec x | m_0, m_{1/2}) \, \pi(m_0,m_{1/2}),
\end{equation}
is shown as a function of $m_0$ and $m_{1/2}$. We consider the set of measured
electroweak observables, $g-2$, $BR(b\to s\gamma)$, $BR(B \to \tau
\nu)$, $BR(B \to D \tau \nu)/BR(B \to D e \nu)$, $R_{{\ell}23}$,
$D_s\to \tau \nu$, $D_s\to \mu \nu$ and $\Delta
\rho$, for which the likelihood is
\begin{equation}
p(\vec X|m_0, m_{1/2}) \propto \prod_i \textrm{Gaussian}(X_i | \alpha_i, \sigma_i),
\end{equation}
where $x_i = \alpha_i(m_0, m_{1/2})$ is the predicted value of the observable $i$ for the model ($m_0$, $m_{1/2}$), which is computed for each observable above using {\tt SuperIso}~\cite{superiso} and {\tt micrOMEGAs 2.4}~\cite{micromegas} and $X_i \pm \sigma_i$ is the associated experimental measurement, in which the central value $X_i$ is taken as the prediction for our TSN, and the uncertainty $\sigma_i$ is taken from the actual measurements quoted by the Particle Data Group~\cite{pdg}.

The plots in Fig.~\ref{fig:posteriorLM1} show that the electroweak results are helpful in
breaking the model degeneracy. We expect this conclusion to remain true for realistic analyses and models.

\subsection{5-D Models}
\label{sec:5DExample}
We now consider two 5-parameter models that illustrate the 
more realistic situation in which the use of a regular grid of parameter
points in the space $\Theta$ rapidly becomes unfeasible due to the well-known ``curse of dimensionality".  
The standard way to circumvent this problem is to sample points using Markov Chain
Monte Carlo. This is what we propose to do in order to approximate the posterior density
$p(\theta)$ where, now, $\theta$ represents a parameter point in the 5-dimensional model
space. 

\subsubsection{Models}
We define two non-universal extensions of the CMSSM that we call NUm$_0$ and NUm$_{1/2}$, which respectively have non-universal $m_0$ and non-universal $m_{1/2}$.  We choose our TSN from NUm$_0$, and therefore also refer to it as the ``TSN model".  We refer to the other model as the ``wrong model". Note that this model cannot be used to parametrize the TSN point due to its universal $m_0$.  The free parameters of the two models and the parameter values at TSN are as follows:

\begin{itemize}
\item TSN model: NUm$_0$ (CMSSM with non-universal m$_{0}$): 
\begin{itemize}
\item $m_0(1,2)$ :  250 GeV at TSN 
\item $m_0(3) = m_{H_{u,d}}$ : 1.5 TeV at TSN
\item $m_{1/2}$ where $m_{1/2} = m_{1/2}(1,2) = m_{1/2}(3)$ : 300 GeV at TSN
\item $A_0$ : 0 GeV at TSN
\item $\tan\beta$ : 10 at TSN
\end{itemize}
\item Wrong model: NUm$_{1/2}$ (CMSSM with non-universal m$_{1/2}$): 
\begin{itemize}
\item $m_0$ where $m_0 = m_0(1,2) = m_0(3) = m_{H_{u,d}}$ 
\item $m_{1/2}(1,2)$
\item $m_{1/2}(3)$ 
\item $A_0$
\item $\tan\beta$
\end{itemize}
\end{itemize}
For both cases, we take the sign of $\mu$ to be positive. 

\subsubsection{Priors}
Our method follows the common Bayesian strategy of ``sacrificing" a small fraction of the data to generate what we have referred to as an \emph{initiating} prior, that is, a prior that  permits the inference chain to proceed. In this example, the multi-parameter priors for the TSN and wrong models are constructed assuming a 100 pb$^{-1}$ data-set.  We again use the {\tt SOFTSUSY}~\cite{softsusy}, {\tt SUSYHIT}~\cite{susyhit}, {\tt PYTHIA}~\cite{Pythia} sequence to generate
events, but {\tt Delphes}~\cite{Delphes} to simulate the CMS detector~\cite{CMSdetector},
and we apply the same CMS jets plus \met\ analysis~\cite{CMSmultijetsMET}. For simplicity, we assume that the subsequent analysis is again that of a counting experiment identical to the one used to construct the priors, except that the integrated luminosity is larger. In practice, one would  work hard to adapt, improve, and change the analyses as more and more data are accumulated.  However, our purpose here is not to do a realistic analysis but simply to illustrate our method. 

\begin{figure}[hbtp]
  \begin{center}
    \includegraphics[width=0.4\textwidth]{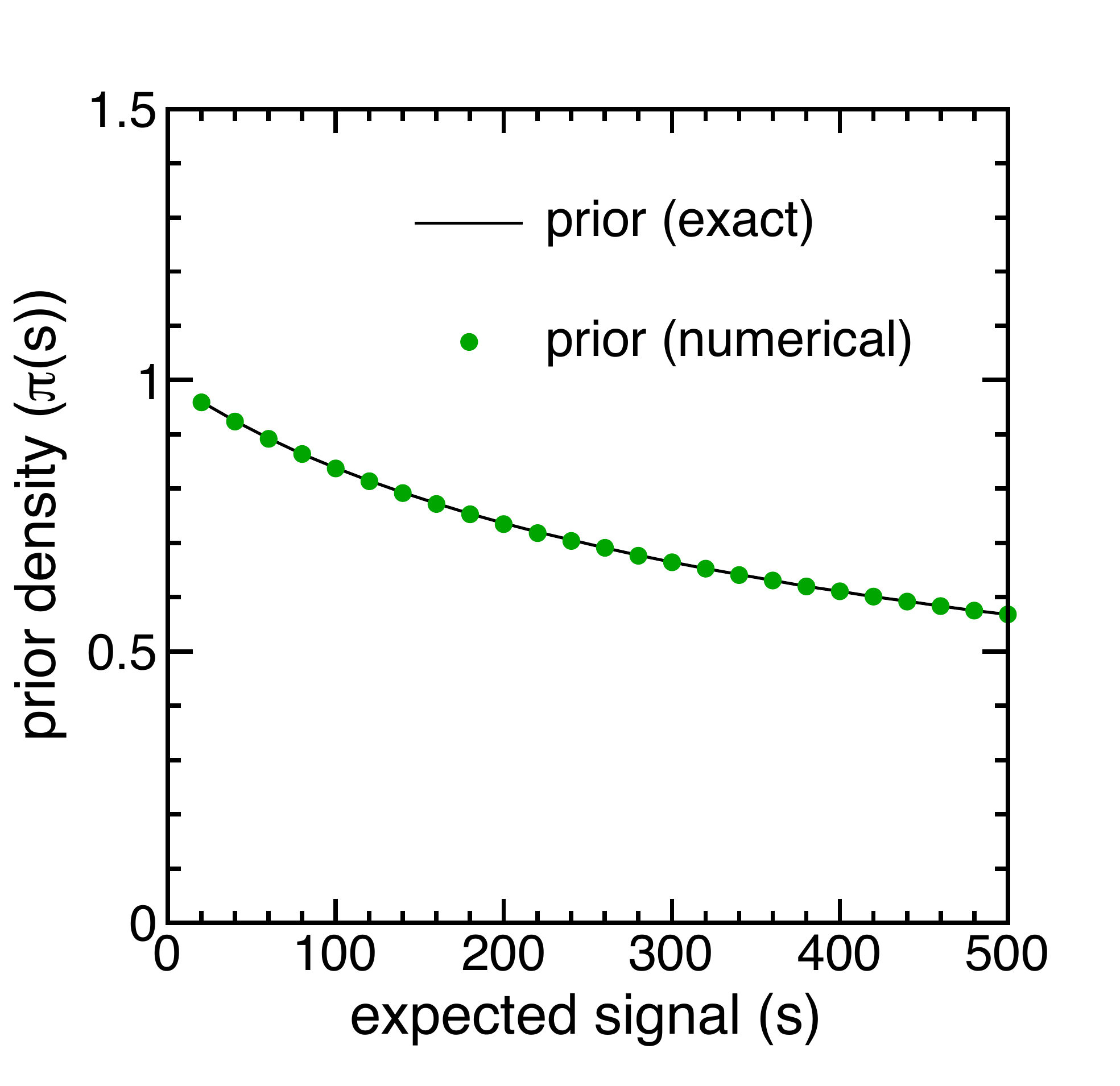}
    \caption{The reference prior, $\pi(s)$, for the single count model computed using \Eq{piR2p} (line) compared
    with  the same computed numerically using \Eq{JeffreysPrior} (points). \label{fig:refprior}}
   \end{center}
\end{figure}

The quantities pertaining to the TSN point, and assuming 100 pb$^{-1}$ are:
\begin{eqnarray}
	\textrm{cross section} \quad \sigma & = & 1.35 \; \textrm{pb}, \nonumber \\ 
	\textrm{signal efficiency}\quad \epsilon & = & 0.412, \nonumber \\
	\textrm{``observed" count} \quad N & = & 169 \quad \textrm{events}, \nonumber \\
	\textrm{background estimate}\quad \hat{\mu} & = & 113 \pm 11.3 \quad\textrm{events}, \nonumber\\
	\textrm{sideband yield}\quad Y & = & 100 \quad\textrm{events}, \nonumber \\
	\textrm{sideband/signal region scale factor} \quad b & = & 0.889. 
\end{eqnarray}
The reference prior using the above values for $Y$ and $b$ is shown in Fig.~\ref{fig:refprior}.
The reference posterior density $p(s|N)$ is computed using the numbers at the TSN point. However, since it is no longer realistic to use a uniform grid of points, we generate a sample of points $\theta_i$ from the reference posterior density $p(s|N)$ with $s = f(\theta)$, for each model, using the Metropolis-Hastings algorithm~\cite{metropolishastings} and multiple MCMC chains. 
Asymptotically,  this sampling procedure will produce a density that satisfies \Eq{intmap}. Moreover, to the degree that the chains can thoroughly explore the surfaces $s - f(\theta) = 0$, the generated points will also satisfy \Eq{ptheta}; that is, the surface term will be automatically incorporated.  The mapping from one  to multiple dimensions is
discussed further in Appendix~\ref{app:2dmapping} using a 2-dimensional toy model. 

\begin{figure}[hbtp]
  \begin{center}
    \includegraphics[width=1\textwidth]{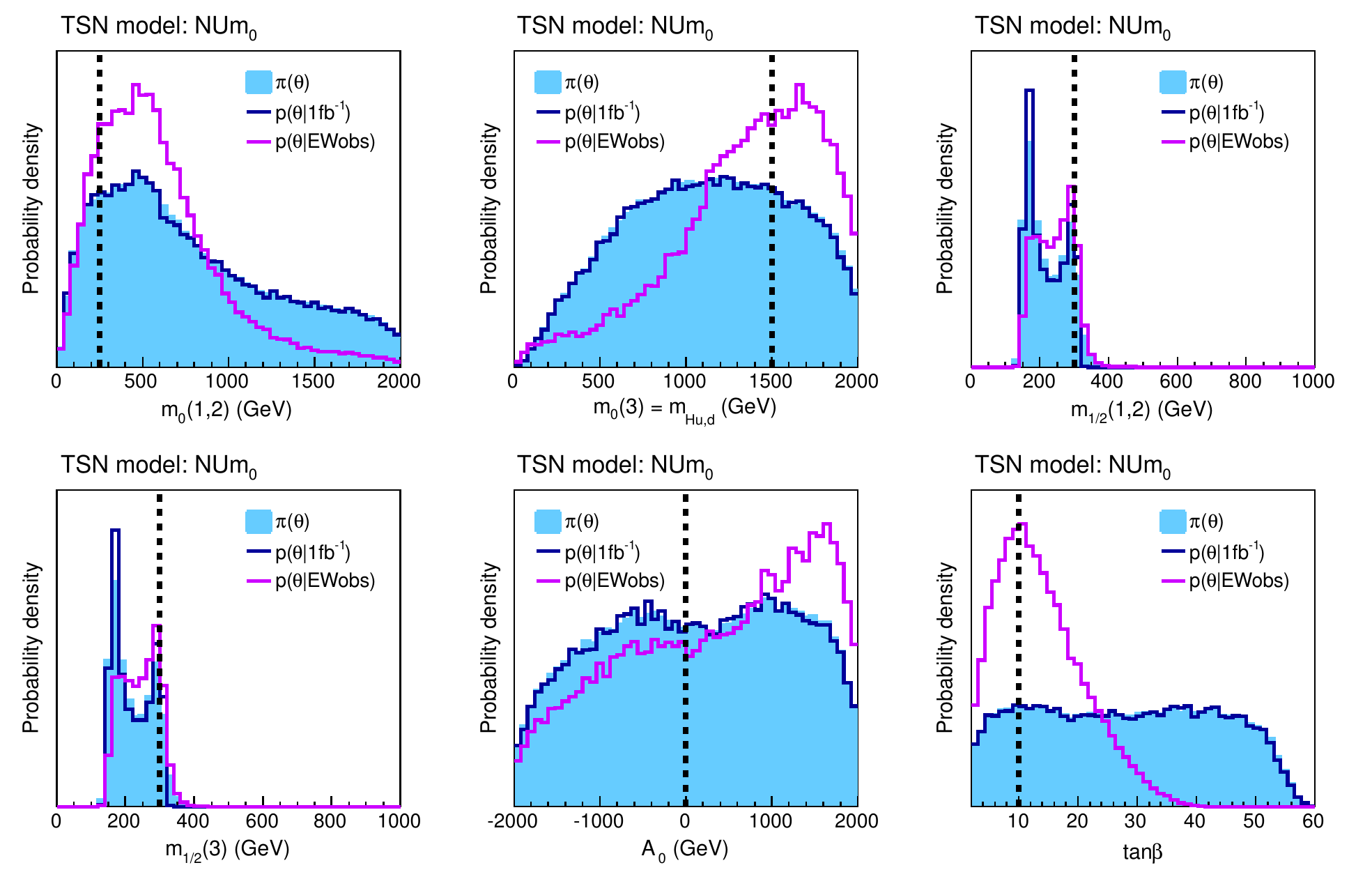}
    \caption{Induced marginal densities for the TSN model assuming a 100 pb$^{-1}$. The
    shaded histograms are the priors. The posterior densities, obtained by weighting the sampled points by the likelihood for  the counting experiment (dark line) and the combined likelihood for the electroweak experiments (light line),  are superimposed on the priors. The vertical dashed line indicates the position of the TSN point. From these projections, one would conclude that the influence of the result of the counting experiment is negligible, while the influence of the electroweak results is quite evident. 
    \label{fig:priorsLM1TSN}}
   \end{center}
\end{figure}

\begin{figure}[hbtp]
  \begin{center}
    \includegraphics[width=1\textwidth]{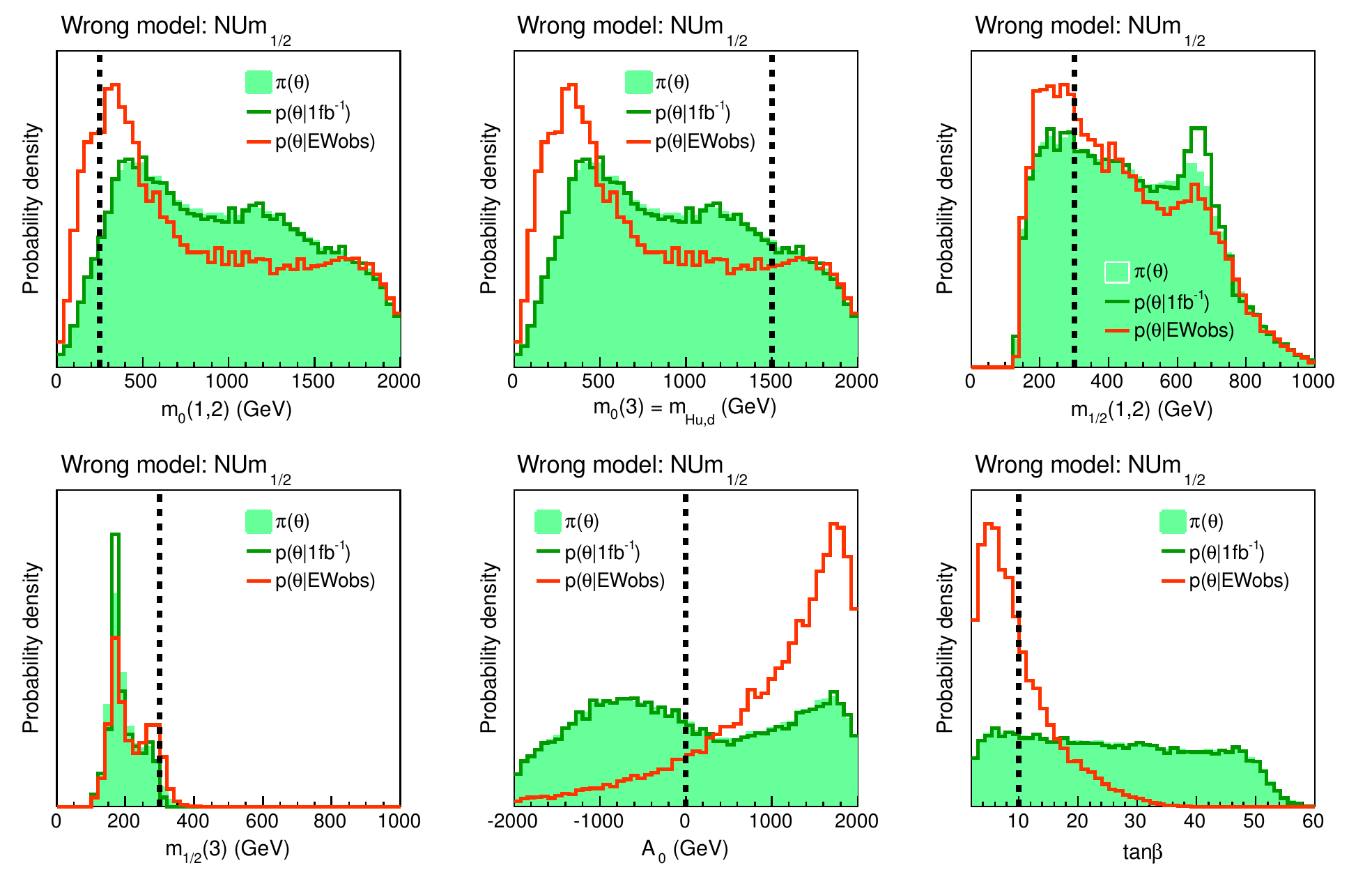}
    \caption{Induced marginal densities for the wrong model. See Fig.~\ref{fig:priorsLM1TSN} for details.  \label{fig:priorsLM1wrong}}
   \end{center}
\end{figure}

Figure~\ref{fig:priorsLM1TSN} shows the 1-dimensional
marginal densities of the induced prior for the TSN model on which are superimposed the posterior densities. The 1-dimensional marginals for the wrong model are shown in Fig.~\ref{fig:priorsLM1wrong}. In both figures, the location of the TSN point  is indicated by the vertical dashed line. Note that in each figure two of the plots are degenerate: the $m_{1/2}(1,2)$ and $m_{1/2}(3)$ plots in Fig.~\ref{fig:priorsLM1TSN} for the TSN model and 
the  $m_0(1,2)$ and $m_0(3)$ plots in Fig.~\ref{fig:priorsLM1wrong} for the wrong model. For the TSN model, most of  the peaks of the 1-dimensional densities are near the TSN point, while for the wrong model this is not the case. 

\begin{figure}[hbtp]
  \begin{center}
    \includegraphics[width=1\textwidth]{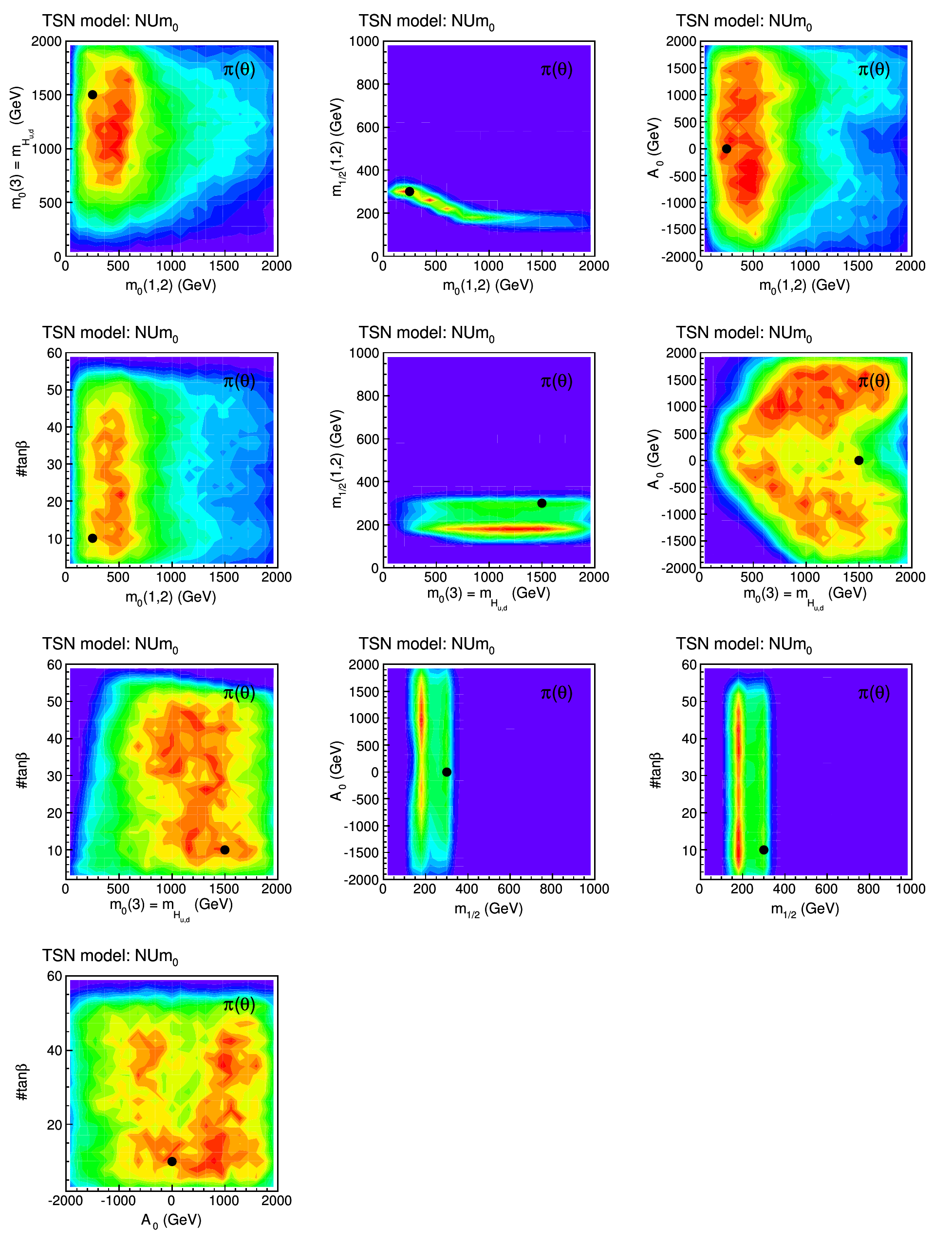}
    \caption{Induced 2-dimensional marginal posterior densities for the TSN model. The TSN is indicated by
    the black dot. See text for details. \label{fig:priorsLM1TSN2d}}
   \end{center}
\end{figure}

We can get a better idea of the shape of the posterior densities from their 2-dimensional marginals, which are shown in Fig.~\ref{fig:priorsLM1TSN2d}. The black point in each plot is the TSN point. One feature which seems puzzling at first is that the TSN point does not always lie at the peak of the densities. But, the following should be noted. If the hyper-surface $s - f(\theta) = 0$ on which the TSN point lies is larger than that of another hyper-surface associated with a smaller value of the reference posterior density $p(s|N)$, then it could happen that the value of $p(\theta)$ on the TSN hyper-surface is actually smaller than its value on the other hyper-surface, even though the total probability of  the TSN hyper-surface is greater than the total probability of other hyper-surfaces. 

\begin{figure}[hbtp]
  \begin{center}
    \includegraphics[width=1\textwidth]{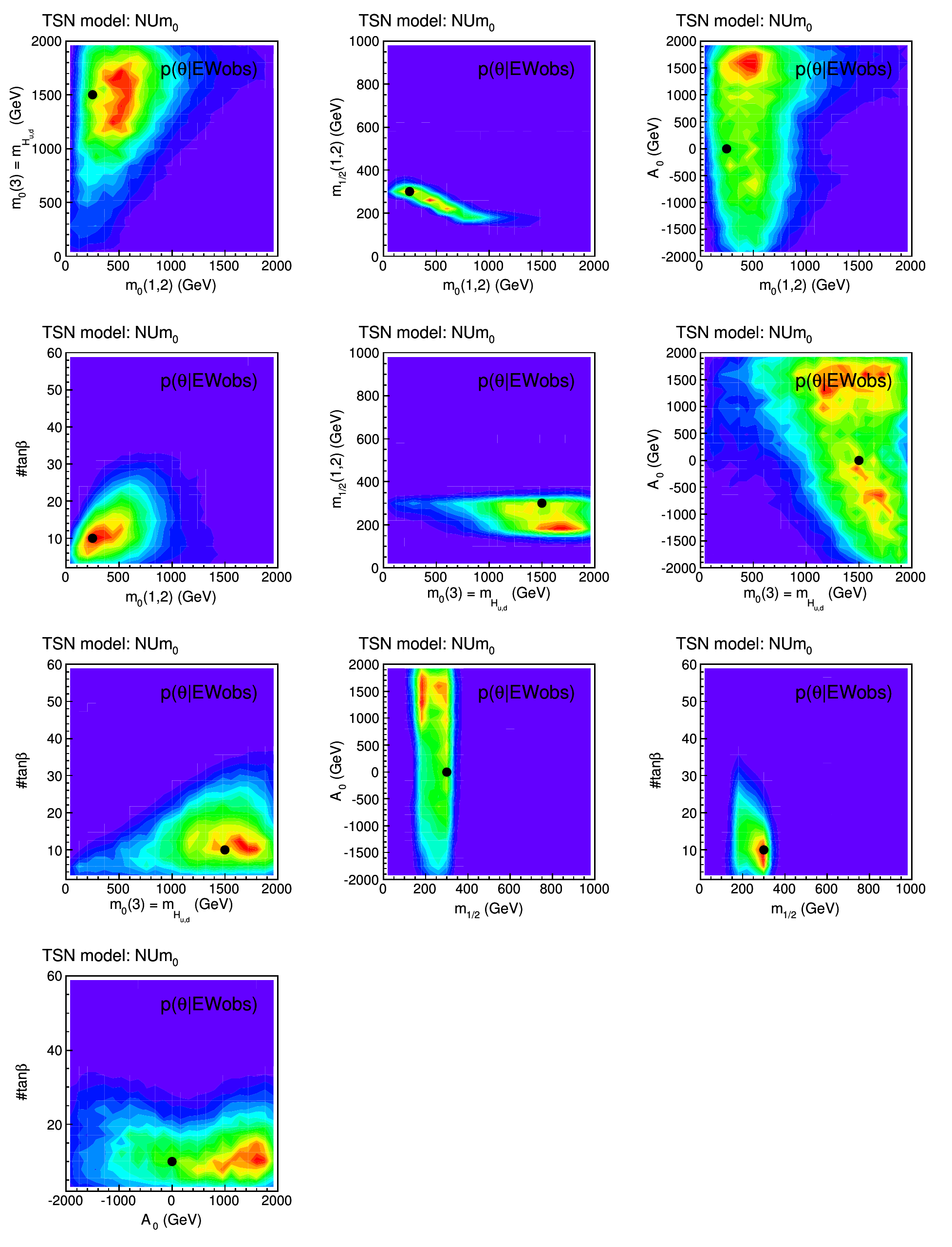}
    \caption{Induced 2-dimensional marginal posterior densities for the TSN model including the effect of
    the electroweak results. The TSN is indicated by
    the black dot. See text for details. \label{fig:priorsLM1TSN2dEW}}
   \end{center}
\end{figure}

Figure~\ref{fig:priorsLM1TSN2dEW} shows what happens to the prior after multiplication by the likelihood for the electroweak results. As expected, these results make a noticeable change to the prior in
sharp contrast to the result of the counting experiment. This is, perhaps, not surprising since the observed count constrains only the signal strength, whereas the electroweak results constrain multiple observables that help break the model degeneracy.

\subsubsection{Signal Significance}
Table~\ref{tab:significance} shows how the signal significance, as defined in \Eq{dN}, increases as
a function of integrated luminosity. We expect this number to scale like $\sim \sqrt{\cal L}$, which
indeed it does.

\begin{table}[htdp]
\caption{Signal significance as a function of integrated luminosity for the TSN model.}
\begin{center}
\begin{tabular}{|c|c|c|c|}
\hline
Integrated luminosity & ``Observed" count (TSN) & \multicolumn{2}{|c|}{Significance} \\
\cline{3-4}
(fb$^{-1}$)	& $N$ events	& $d(N)$	& $\sqrt{2 \, d(N)}$ \\
\hline
0.5 	& 331 	& 12.2 	& 4.9		\\
1.0 	& 387 	& 13.6 	& 5.2		\\
2.0 	& 660 	& 19.2 	& 6.2		\\
5.0 	& 1754 	& 33.2 	& 8.2 	\\
\hline
\end{tabular}
\end{center}
\label{tab:significance}
\end{table}

\subsubsection{Model Ranking}
As we noted, the purpose of this example is to illustrate the prior construction method. However, it is interesting to
see what happens if we try to rank the TSN and wrong models on the basis of the signal
strength only. The results are shown in Table~\ref{tab:ranking}. We find that even with the relatively weak constraint afforded by merely counting events, we are able to rank these models consistently, albeit weakly.

\begin{table}[htdp]
\caption{Ranking of the TSN and wrong models as a function of integrated luminosity.}
\begin{center}
\begin{tabular}{|c|c|c|c|}
\hline
Integrated & Evidence for & Evidence for & Evidence TSN over \\
luminosity & TSN model     & wrong model & Evidence wrong model \\
\hline
0.5 fb$^{-1}$ & 0.00253 & 0.00205 & 1.233 	\\
1.0 fb$^{-1}$ & 0.00203 & 0.00164 & 1.235	\\
2.0 fb$^{-1}$ & 0.00102 & 0.00083 & 1.238	\\
5.0 fb$^{-1}$ & 0.00034 & 0.00028 & 1.245	\\ 
\hline
\end{tabular}
\end{center}
\label{tab:ranking}
\end{table}

\section{Summary and Conclusions}
We have proposed a method for building multi-parameter priors that follows the general
strategy of building a proper prior using a small portion of the data and analyzing the rest using that
prior. Since the direct construction of multi-parameter priors, with
mathematically well-defined properties, is a difficult task we have proposed a method that
begins with a simpler task, namely, the construction of a reference prior for an analysis having
a single parameter of interest. Together with the likelihood function, the reference prior yields a proper posterior density that is consistent with a class of posterior densities on the parameter space of the physics
model under study. We proposed choosing a particular member from this class to serve as the multi-parameter prior for subsequent analyses. That prior has the property that its
density is constant on every hyper-surface indexed by the parameter of interest. Moreover, because
it is built from a reference prior, the multi-parameter prior is expected to yield
credible regions with excellent frequentist
properties. Finally, the robustness of inferences can be assessed by weighting the multi-parameter prior $\pi(\theta)$ by, for example, $w(s) = [A(s)/p(s|N)]^r$ and studying  the sensitivity
of inferences to the exponent
$0 \leq r \leq 1$. The exponent $r$ permits a smooth interpolation between the reference prior $(r = 0)$ and
a flat prior ($r = 1$).

Our proposed construction must surmount a technical hurdle: generating a sample of points in the parameter space of the physics model with the properties that 1) the number of points  on 
each hyper-surface is
proportional to the reference posterior density associated with that hyper-surface and 2) the points 
on the hyper-surface are uniformly
distributed. We showed, using three illustrative examples,  how one might address this question, in general. For high-dimensional models, the use of MCMC seems feasible. However, we have found that convergence may be an issue because of the severe degeneracies  present when relatively little information is used to create the multi-parameter prior. In a realistic application it will be necessary to tune the MCMC algorithm to ensure convergence of the Markov chains. It would be useful to explore different sampling methods,  such as {\tt MultiNest}~\cite{multinest}, that may be better suited to problems with severe degeneracies. 

In spite of these challenges, however, we have shown that our method yields priors that give consistent results as
more and more data are accumulated. What remains to be done is to apply the method to a real analysis at the LHC. Our expectation is that the method would fare well.

\begin{acknowledgments}
We thank Jim Berger and Jos\'{e} Bernardo
for discussions on reference priors and Bayesian methods in general and
Sabine Kraml for discussions on the SUSY models. We also thank Luc Demortier, Bob Cousins,
and Kyle Cranmer for several discussions that helped clarify our thoughts.

This work was supported in part by the U.S. Department of Energy under grant  no. DE-FG02-97ER41022.
\end{acknowledgments}



\appendix
\section{Derivation of Background Prior}
\label{app:backgroundPrior}
This form for the prior $\pi(\mu)$ can be motivated~\cite{Demortier2010} by considering an
experiment comprising two data-sets $S$ and $B$. Data-set $S$ is modeled as
a mixture of signal and background events
with expected background count $\mu$. Data-set $B$, perhaps a sideband, is presumed to be overwhelmingly dominated by background
events with expected background
 $b\mu$. Although we do not know $\mu$,
we assume that we know the ratio $b$ of the expected background in data-set $B$ to
that in data-set $S$. The expected background $b\mu$ for
data-set $B$ is estimated by the number of events $Y$
in that data-set. The likelihood for the observed count $Y$ in data-set $B$ 
is taken to be $\mbox{Poisson}(Y | b\mu)$, which, together
with its
reference prior, $\propto 1/\sqrt{\mu}$, yields the posterior density 
$p(\mu | Y) \propto \exp(-b\mu) (b\mu)^{Y-1/2}$. This posterior density serves as the evidence-based prior $\pi(\mu)$ for
the expected background in data-set $S$.

\section{Definition of Reference Prior for the Single Count Model}
\label{app:refPrior}

One begins with the information gained from
$K$ repetitions of the single count experiment, 
\begin{equation}
I_K[\pi] \equiv \sum_{N_1 = 0}^{\infty} \cdots \sum_{N_K = 0}^{\infty}  m(N_{(K)}) \, D [\pi , p(s | N_{(K)} )],
\label{eq:Ik}
\end{equation}
 where  
\begin{eqnarray}
m(N_{(K)}) \; & =  & \; \int p(N_{(K)} | s ) \, \pi(s) \, ds, \nonumber \\ 
\mbox{with} \; p(N_{(K)} | s) \; & = & \; \prod_{i=1}^{K} p(N_i | s), 
\end{eqnarray}
is the marginal density for
$K$ experiments. The maximization of the expected information gain, $I_K[\pi]$, with respect to the 
prior yields the function $\pi_K(s)$. 
By definition~\cite{Berger2009}, the reference prior $\pi(s)$ is the limit
\begin{eqnarray}
\pi(s) \; & =  & \; \lim_{K\rightarrow\infty}
                       \frac{\pi_{K}(s)}{\pi_{K}(s_{0})}, \nonumber \\
\mbox{with} \; \pi_K(s)   \; &  =  & \; \exp\left\{ \sum_{N_1 = 0}^{\infty} \cdots \sum_{N_K = 0}^{\infty}   p(N_{(K)} | s)\,
                \ln\left[\frac{p(N_{(K)} | s)\, h(s)}
                {\int  p(N_{(K)} | s) \, h(s)\, ds}\right]\,\right\},
\label{eq:refprior}
\end{eqnarray}
where $s_{0}$ is any fixed point in the space of expected signal and $h(s)$ is any positive function, such as $h(s)=1$. However, since the posterior density for the single count model is asymptotically normal, the reference prior computed using the above algorithm coincides 
with Jeffreys prior, Eq.~(\ref{eq:JeffreysPrior}).

\section{Calculation of Marginal Likelihood}
\label{app:likelihood}

Defining the recursive functions,
\begin{equation}
\begin{split}
W_0(s, z)   & \;=\; 1, \\[2mm]
W_{k}(s, z) & \;=\; z \left( \frac{s}{k} \right)\;W_{k-1}
                    \quad\textrm{for}\quad k=1,\cdots,n,\\[2mm]
Y_0(z)      & \;=\; 1, \\[2mm]
Y_k(z)      & \;=\; z \left( \frac{y-\frac{1}{2}+k}{k} \right) 
                    \left( \frac{1}{b+1} \right)\;Y_{k-1},
                    \quad\textrm{for}\quad k=1,\cdots,n,
\end{split}
\label{eq:WXY}
\end{equation}
we can write $p(n\,|\,s)$ and $T_n^m(s)$ as
\begin{eqnarray}
 p(n\,|\,s) &\;=\;&\left[\frac{b}{b+1}\right]^{y+\frac{1}{2}}\;
                  \sum_{k=0}^{n} W_k(s, z) \; Y_{n-k}(z),\\
  T_n^m &\;=\;&\sum_{k=0}^n k^m \; W_k(s, z) \; Y_{n-k}(z),       
                  \label{eq:MarginalModelCalc}
\end{eqnarray}
with $z=1$ for $n = 0$ and $z= e^{-s/n}$ for $n > 0$.

%
%
%
%
%

\section{Mapping Procedure for a 2D Toy Model}
\label{app:2dmapping}
To illustrate further how the mapping from a 1-D posterior density to an $n$-D parameter
space works in practice, we consider the case of a model described
by two unknown parameters $x$ and $y$. An experimental measurement
is available for the quantity $\rho = \sqrt{x^2+y^2}$. One builds the 
reference prior corresponding to all the possible outcomes of the
measurement of $\rho$ and derives a reference posterior $p(\rho)$.
We now want to find a function $\pi(x, y)$ that is consistent
with the 1-D reference posterior density $p(\rho)$. 

To solve this problem, we impose two
conditions:
\begin{itemize}
\item  $\pi(x,y)$ is constant for all the points $(x,y)$
  corresponding to the same value of $\rho$. This implies that
  $\pi(x,y)= \pi(\rho(x,y))$. This makes perfect sense because the
  only available information on $x$ and $y$ is the measurement of
  $\rho$, which cannot break the degeneracy of the iso-$\rho$ contour.
  Without any loss of generality, we can then write $\pi(x,y)=p(\rho(x, y)) / A(\rho(x, y))$;
\item when marginalized to $\rho$,  through \Eq{intmap}, $\pi(x,y)$ should recover
  $p(\rho)$. This consistency requirement, together with the first, 
  is what permits identifying
  $A(\rho(x,y))$ with the ``area" of the iso-$\rho$ contour. 
\end{itemize}
The first requirement is quite natural if one thinks of the Bayesian
analysis as an update of our knowledge about the parameters $x$ and $y$. 
The second
requirement may need further explanation.

\begin{figure}
\centering\includegraphics[width=.45\linewidth]{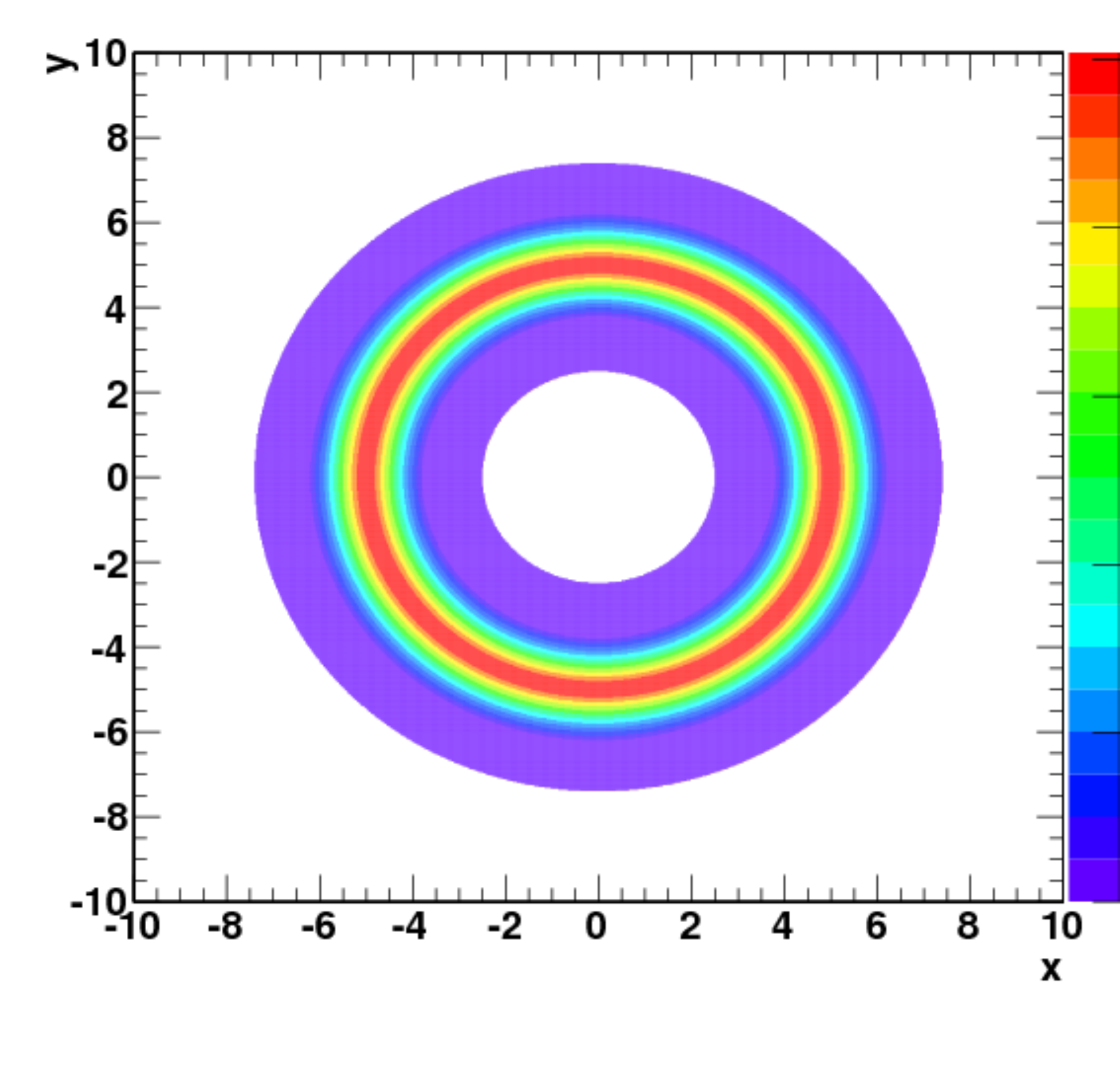}
\centering\includegraphics[width=.45\linewidth]{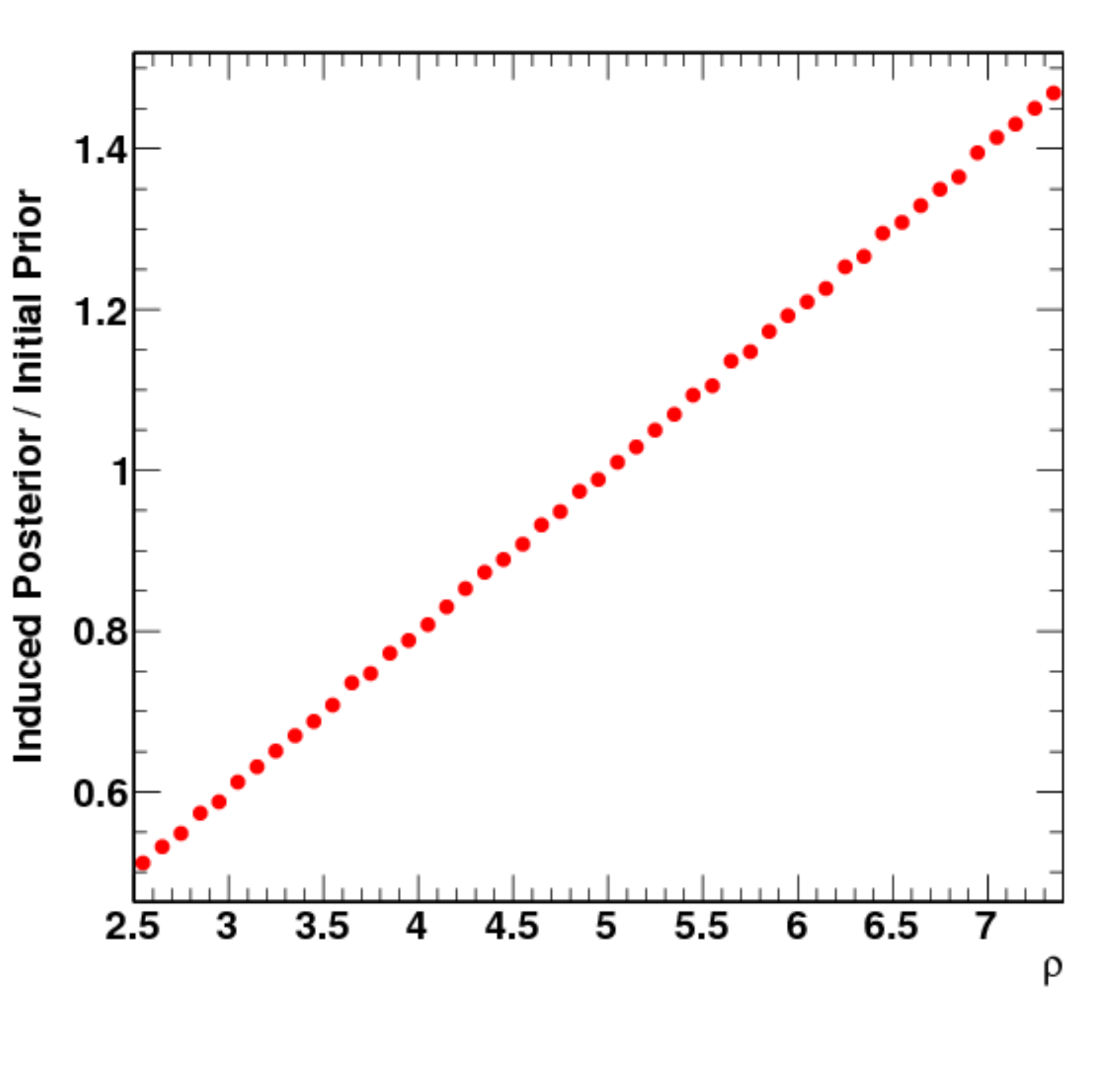}
\caption{(left) Induced posterior density $p^\prime(x, y) = p(\rho(x, y))$, where $p(\rho)$ is the
1-D reference
posterior density. (right) Ratio of $p^\prime(x,y)$ marginalized back to $\rho$, via
\Eq{intmap}, over the
reference posterior density $p(\rho)$. Clearly the two 1-D densities are not the same, as
they should be if the density $p^\prime(x, y)$ were consistent with $p(\rho)$.
\label{fig:wrongMap}}
\end{figure}

Suppose for the moment that the function $A(\rho(x,y))$ does not
enter the problem. Enforcing the first condition would then imply that
$\pi(x,y)=p(\rho(x,y))$. Consider a measurement of $\rho$
with a Gaussian likelihood. This measurement would translate
into a 2-D function of $x$ and $y$ as shown in the left plot of 
Fig.~\ref{fig:wrongMap}.
Once marginalized, this function gives a function $g(\rho)$
which differs from $p(\rho)$ by a factor linear in $\rho$, coming
from the Jacobian of the $(x, y) \to \rho$ marginalization. This is
shown in the right plot of Fig.~\ref{fig:wrongMap}, which shows
the ratio
$g(\rho) / p(\rho)$ as a function of $\rho$. 

\begin{figure}
\centering\includegraphics[width=.45\linewidth]{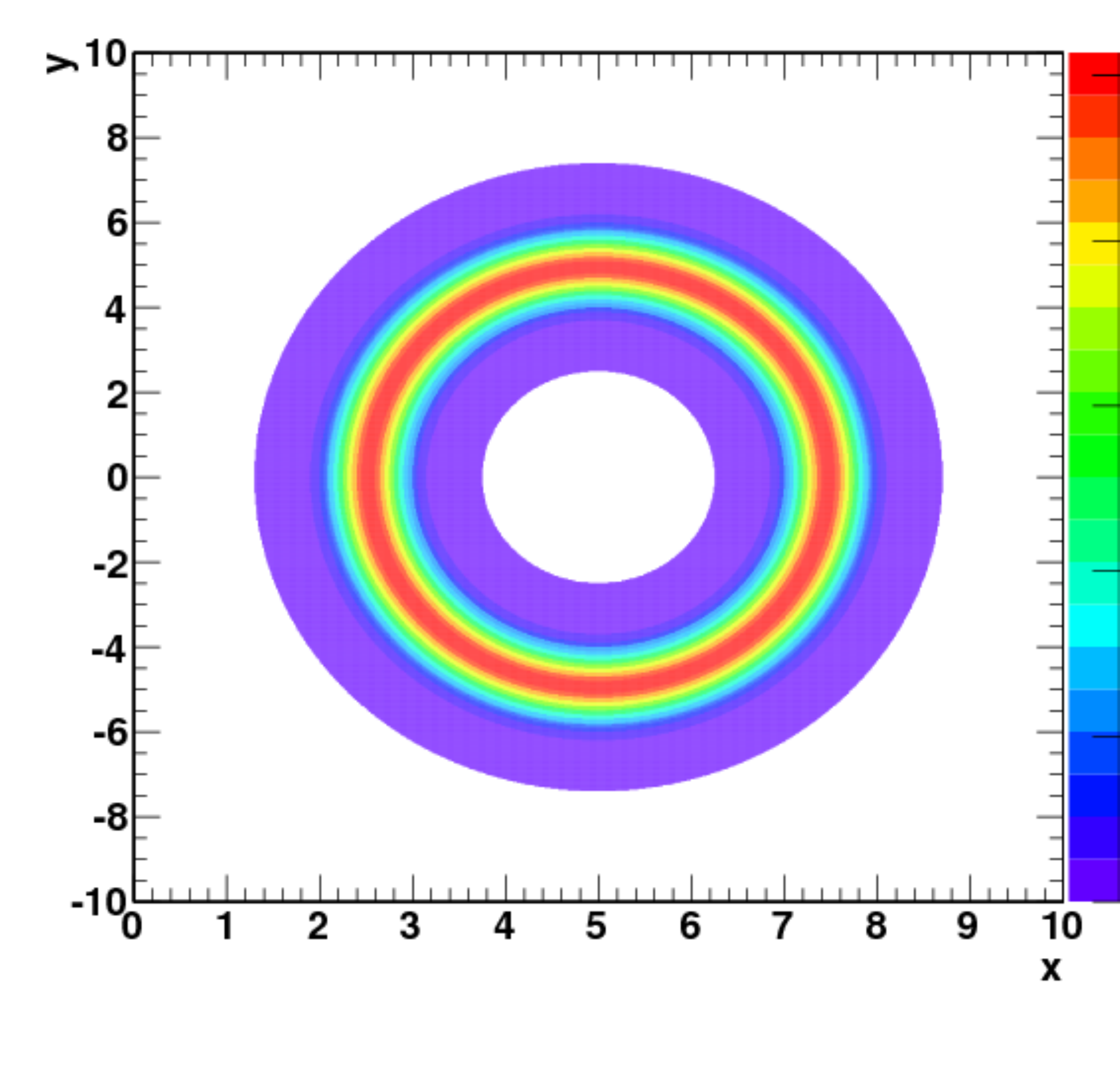}
\centering\includegraphics[width=.45\linewidth]{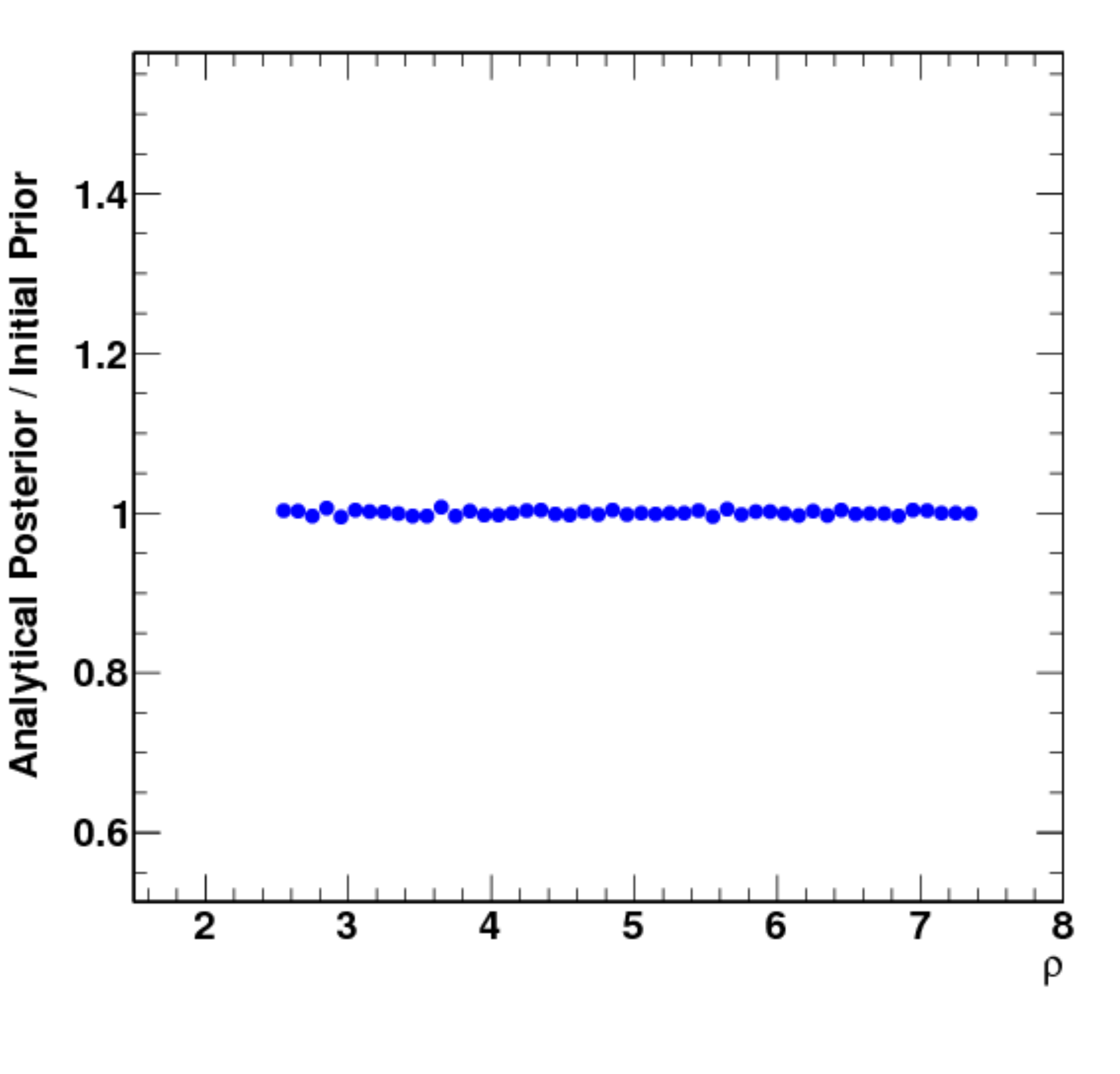}
\caption{(left) Induced posterior density $p^\prime(x, y) = p(\rho(x, y)) / 2\pi \rho$, where $p(\rho)$ is the
1-D reference
posterior density. (right) Ratio of $p^\prime(x,y)$ marginalized back to $\rho$, via
\Eq{intmap}, over the
reference posterior density $p(\rho)$.  The two 1-D densities are identical, as
they should be since, by construction, the density $p^\prime(x, y)$ is consistent with $p(\rho)$.\label{fig:anMap}}
\end{figure}

However, in this specific case, we know the form of the function $A(\rho)$; it 
is simply given by
$A(\rho)=2\pi\rho$. Therefore, the correct mapping from 1-D to 2-D yields
 $\pi(x,y) = p(\rho(x, y)) / 2\pi\rho$, 
 shown in the left plot of Fig.~\ref{fig:anMap},
which gives a constant value for the ratio $g(\rho) / p(\rho)$ (see right plot of Fig.~\ref{fig:anMap})
as one would expect for a density $\pi(x, y)$ that is consistent with $p(\rho)$.

\begin{figure}
\centering\includegraphics[width=.45\linewidth]{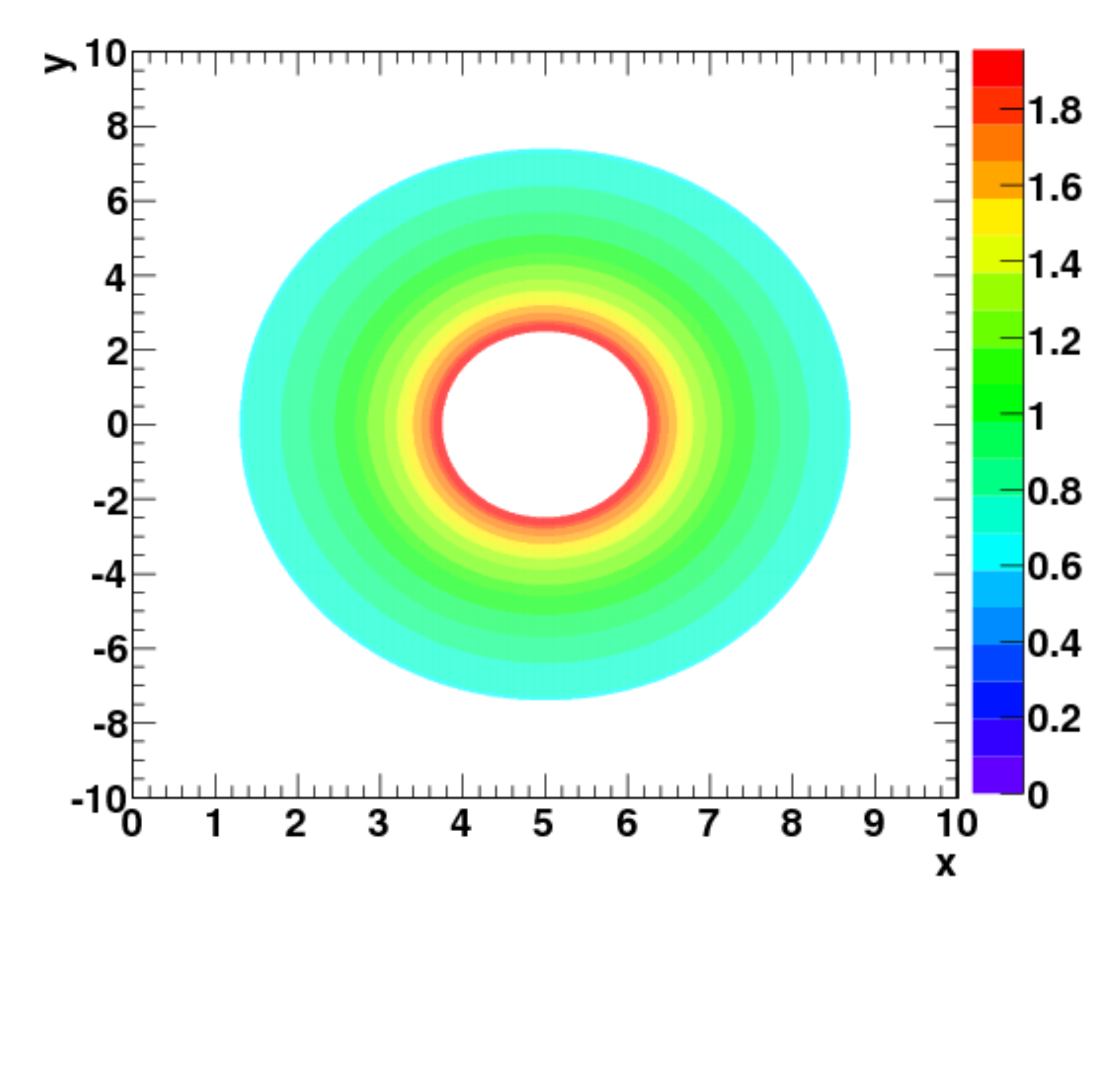}
\centering\includegraphics[width=.45\linewidth]{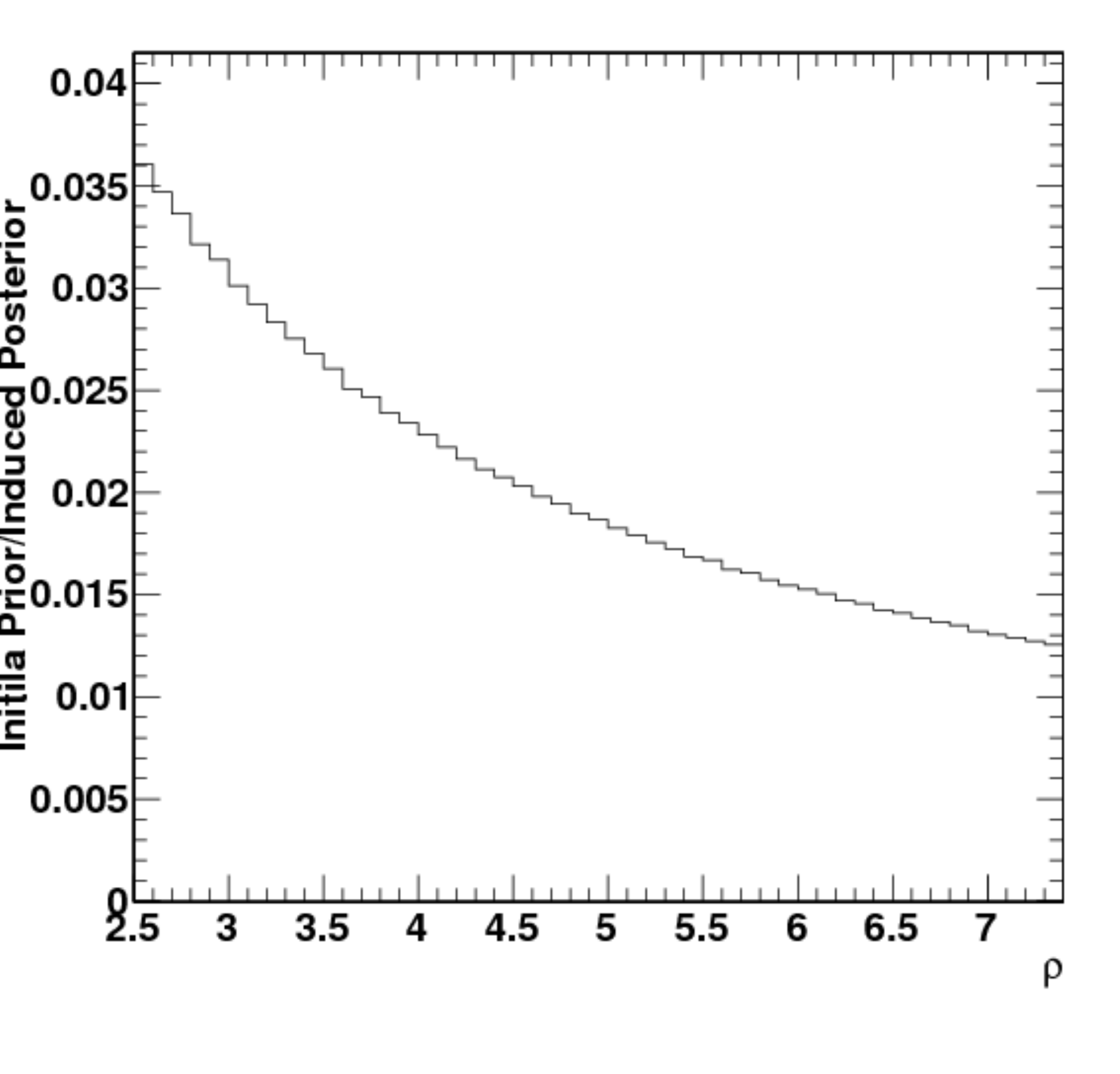}
\caption{(left) Correction map in the $x, y$ plane and (right) the same map
in the $\rho$ space. \label{fig:correctedMap}}
\end{figure}

In the absence of an analytical solution for $A(x, y)$, one could follow a
simple numerical procedure, which takes full advantage of the fact
that $A(x,y) = A(\rho(x,y))$. This simple fact implies that, by incorrectly
using $p^\prime(x,y) = p(\rho(x,y))$ one is wrong by a factor that is
constant over the iso-$\rho$ contour. This factor is nothing else than
the ratio $g(\rho) / p(\rho)$, mapped onto the $(x,y)$ plane (see left
plot of Fig.~\ref{fig:correctedMap}). This simple construction allows one to solve for the
integral, \Eq{Area}, defining $A(x,y)$ without having to perform the integral explicitly; one
simply weights each point by $g(\rho) / p(\rho)$,  which is shown in the right-hand plot
of Fig.~\ref{fig:correctedMap}). When the 
corrected function $\pi(x, y)$ is marginalized, the function $p(\rho)$ is recovered by
construction.

The use of MCMC to sample the space $x, y$ makes the procedure even simpler.
Rather than scanning the $(x,y)$ plane and associating to each point the 
value of $p(\rho)$, one samples  $(x,y)$ according to $p(\rho)$
directly. This implies that $g(\rho) = p(\rho)$ by construction, as one
can easily verify.

\end{document}